# Advancements in Cloud-Native Systems: A Systematic Mapping Study of Reliability and Scalability across Cloud-Native, Edge-Native and AI-Enabled Systems

SHRINIVASS ARUNACHALAM BALASUBRAMANIAN

Independent Researcher, Shrewsbury, MA, USA

Corresponding author: Shrinivass Arunachalam Balasubramanian, shrinivassab@gmail.com

Abstract. Cloud-native systems now extend to the network edge and host artificial-intelligence (AI) workloads, including AI agents, but evidence on their reliability and scalability is scattered and uneven. We report a PRISMA 2020 systematic mapping study with an evidence-graded synthesis. Five logged OpenAlex searches on 11 October 2026 covered cloud-native systems, edge AI, agentic AI for operations, reliability of agentic applications and physical AI; 189 records from an earlier exploratory search were added separately. After 301 duplicate removals and an automated peer-review venue rule, 4,611 records were screened against written criteria and a five-level relevance scale. The review includes 1,794 primary studies from 2017–2026 and 95 secondary studies, with every decision released in Dataset S1. Each primary study was assigned to a 13-category taxonomy and graded as simulation, prototype or testbed, production data, production deployment, or not determinable from the abstract. Of 1,003 studies with a determinable level, 66.2% used prototypes or testbeds, 18.0% simulations and 15.8% production data or deployments. A 93-study synthesis core, appraised from indexed abstracts, shows production deployments of cascading-failure prevention, trace- and graph-based root-cause analysis and predictive autoscaling, and production-trace evaluations of cold-start mitigation. Edge reliability and physical AI rely mainly on simulations and testbeds. AI-agent reliability uses tool-grounded diagnosis, explicit safety properties, trace-driven repair and workload-aware scheduling, while empirical studies show that agent frameworks remain fragile. Conclusions are limited to the documented corpus. The principal gaps are production evidence for edge-native systems, shared resilience benchmarks and dependable agentic operations.

## 1 INTRODUCTION

Cloud-native computing has become the default way to build and operate large online services. Applications are decomposed into microservices or serverless functions, packaged in containers, deployed by declarative orchestrators such as Kubernetes and operated through automation and telemetry (Burns et al., 2016; CNCF, 2018). The properties that make these systems easy to scale also make them hard to keep reliable. A failure in one service can cascade through its callers (Chabbi et al., 2026), overload must be controlled across thousands of interdependent services (Zhou et al., 2018b), and diagnosing a fault requires correlating traces, metrics and logs across many components (Zhou et al., 2018a; Li et al., 2021a). Production incidents in large cloud services still take a long time to detect, root-cause and mitigate (Ghosh et al., 2022).

Two developments widen the problem. First, computation is moving towards the edge of the network, close to the devices that produce data and the users who consume results (Shi et al., 2016; Satyanarayanan, 2017). Edge nodes are heterogeneous, resource-constrained and intermittently connected, so applications designed for them, which Satyanarayanan et al. (2019) call edge-native applications, face failure modes that data-centre designs do not assume. Second, AI workloads are increasingly part of these systems. Models run on edge servers and devices (Zhou et al., 2019a); large language model (LLM) agents are used to operate cloud systems (Wang et al., 2024c; Chen et al., 2025); agent applications have themselves become distributed systems whose reliability must be engineered (Xue et al., 2025; Wang et al., 2026b); and robots offload perception and planning to edge and cloud resources (Kehoe et al., 2015; Ichnowski et al., 2023).

Secondary studies cover failure diagnosis, observability, fault injection, cold starts, autoscaling, performance assessment, edge intelligence, TinyML, cloud robotics and agentic AI (Fu et al., 2025a; Gomes et al., 2025; Yu et al., 2025; Golec et al., 2024; Jeong et al., 2025; Rodrigues et al., 2025; Wang et al., 2024b; Schizas et al., 2022; Afrin et al., 2021; Molinari & Ciravegna, 2026). None maps cloud-native, edge-native and AI workloads together while grading each mechanism by evaluation setting. This distinction separates simulation, testbed and production claims.

This paper addresses the gap with a systematic mapping study (Petersen et al., 2015) combined with an evidence-graded narrative synthesis, reported following PRISMA 2020 (Page et al., 2021). It answers four research questions (RQs):

- **RQ1.** Which mechanisms improve reliability and fault tolerance in cloud-native and edge-native systems, including resilience testing and chaos engineering, and in which evaluation settings have they been validated?
- **RQ2.** How do observability, anomaly detection and root-cause analysis support operational resilience, and what evidence supports them?
- **RQ3.** Which scalability and resource-management mechanisms (autoscaling, cold-start mitigation, scheduling, placement and service configuration) have been proposed for microservice, serverless and edge deployments, and which trade-offs do the studies report?
- **RQ4.** How are reliability and scalability ensured for AI at the edge (edge AI and on-device analytics), for agentic AI, both as an operator of cloud systems and as an application in its own right, and for physical AI?

The paper contributes:

- working definitions used throughout (Tables 1 and 2);
- a reproducible search and screening of 7,112 database records plus 189 earlier records, yielding 1,794 primary studies with all decisions in Dataset S1;
- a 13-category, evidence-graded taxonomy showing where production evidence exists (Section 4.3);
- an appraised 93-study synthesis for RQ1–RQ4 (Sections 5–8 and Supplementary Tables S1–S4); and
- a research agenda derived from documented gaps (Section 9.2).

After definitions and related work, the paper describes the method, presents the map, synthesises RQ1–RQ4, discusses the evidence and closes with validity threats and conclusions.

## 2 BACKGROUND AND DEFINITIONS

### 2.1 Cloud-native and edge-native systems

The Cloud Native Computing Foundation defines cloud-native technologies as enabling scalable applications in dynamic clouds through containers, service meshes, microservices, immutable infrastructure and declarative APIs (CNCF, 2018). Kubernetes grew from Google's Borg and Omega experience (Burns et al., 2016). Here, a cloud-native system consists of independently deployable services or functions, managed runtimes, declarative orchestration and elastic data-centre resources. Its reliability model assumes broadly homogeneous capacity and recovery by restart, rescheduling or replication.

Edge computing moves computation close to data sources and users to reduce latency, backhaul use and data movement (Shi et al., 2016; Satyanarayanan, 2017). An edge-native application depends on those properties (Satyanarayanan et al., 2019). We define an edge-native system as one designed for dispersed, constrained and heterogeneous nodes, often with a cloud tier. It treats intermittent connectivity, limited energy and local autonomy as primary constraints. It may reuse cloud-native tools, but capacity cannot always be added and recovery may proceed without the cloud.

We use **edge–cloud continuum** for systems that deliberately span both tiers. When a study deploys cloud-native tooling on edge hardware, we classify it by the problem it addresses (for example, autoscaling or fault tolerance) and record its deployment scope as edge or edge–cloud. Table 1 summarises the distinction. The terms carry these meanings in every later section.

**Table 1.** Working definitions used in this review: cloud-native versus edge-native systems.

Sources: CNCF (2018), Burns et al. (2016), Shi et al. (2016), Satyanarayanan (2017) and Satyanarayanan et al. (2019), complemented by the failure and scalability concerns reported in the reviewed studies.

### 2.2 Reliability, scalability and observability

We follow the dependability terminology of Avizienis et al. (2004). *Reliability* is continuity of correct service, *availability* is readiness for correct service, and *fault tolerance* is the ability to avoid service failures in the presence of faults. We use *resilience* for the ability to keep delivering acceptable service, or to restore it quickly, when faults, overload or changes occur, which is how most reviewed studies use the term. *Scalability* is the ability of a system to sustain growing load by using additional resources, and *elasticity* is the degree to which it adapts its resources to load changes automatically and in time (Herbst et al., 2013). *Observability* is the ability to infer a system's internal state from the telemetry it emits, usually metrics, logs and distributed traces (Li et al., 2021a; Usman et al., 2022).

### 2.3 AI at the edge and AI agents

AI plays several roles, so Table 2 fixes the terminology. Edge AI runs inference, and sometimes training, near the data source (Zhou et al., 2019a); on-device analytics runs on the end device, including TinyML hardware (Schizas et al., 2022). AI-driven edge intelligence manages models, configurations, placement or offloading. Agentic AI uses LLM agents that plan, call tools and act in multi-step loops (Yao et al., 2023; Wang et al., 2024a); we distinguish agents that operate clouds (C11) from agent applications whose own reliability and scalability must be engineered (C12). Physical AI acts through robots or other cyber-physical systems (Duan et al., 2022) and is included when it uses edge or cloud resources (Kehoe et al., 2015).

**Table 2.** Working definitions of the AI-related terms and the taxonomy categories that cover them.

Sources: Zhou et al. (2019a), Schizas et al. (2022), Yao et al. (2023), Wang et al. (2024a), Duan et al. (2022) and Kehoe et al. (2015). Learning-based resource management is classified by the mechanism it implements (for example, a learned autoscaler under C6); when the managed resources are at the edge, these studies are counted as AI-driven edge intelligence in Section 8.1.

### 2.4 Related secondary studies

Supplementary Table S5 lists the secondary studies closest to this review; all 95 secondary studies found by the search are listed in Dataset S1. Most cover one mechanism family for microservice or serverless systems: failure diagnosis and root-cause localisation, observability, fault injection, cold starts, autoscaling, or performance and scalability assessment. Four cover AI-related settings: trustworthy edge intelligence (Wang et al., 2024b), TinyML (Schizas et al., 2022), agentic generative AI in pervasive computing (Molinari & Ciravegna, 2026) and resource allocation in multi-agent cloud robotics (Afrin et al., 2021). As far as their abstracts indicate, none combines cloud-native, edge-native and AI workloads in one taxonomy or grades the primary evidence by the setting in which each mechanism was evaluated. This review adds both, and it releases its complete screening record so that its counts can be verified.

**Supplementary Table S5.** Related secondary studies and their overlap with this review.

## 3 REVIEW METHOD

### 3.1 Design

The review combines a systematic mapping study (Petersen et al., 2015), which classifies all eligible primary studies, with a structured narrative synthesis of a core set appraised from indexed abstracts for each research question. Search, selection and reporting follow the software-engineering guidelines of Kitchenham & Charters (2007) and the PRISMA 2020 statement (Page et al., 2021); Figure 1 gives the PRISMA flow. No review protocol was registered in advance. The review was conducted by a single author with assistance from a large language model (LLM), as described under "Screening procedure and use of AI assistance". Every record, decision and code is available in Dataset S1, which also contains the search log.

### 3.2 Information source and search strategy

We searched OpenAlex (Priem et al., 2022) on 11 October 2026 using five blocks (Supplementary Table S6): cloud-native reliability and scalability (S1), edge AI (S2), agentic AI for operations (S3), reliability of agentic applications (S4) and physical AI across the edge–cloud continuum (S5). Each block combines concept groups with AND and terms within groups with OR. Searches covered 2017–2026, except S3 and S4 (2023–2026), and included articles, conference papers, book chapters and reviews.

**Supplementary Table S6.** Search blocks run in OpenAlex on 11 October 2026.

Document types: article, conference paper, book chapter, review. The exact strings follow in the text; Dataset S1 (sheet Search_log) records the API filters and the hit count at retrieval for each block.

The exact Boolean strings follow. Each block lists its concept groups with the field searched; the groups of a block are combined with AND.

**S1.** Title: ("cloud native" OR microservice OR microservices OR serverless OR "function as a service" OR FaaS OR Kubernetes OR "service mesh" OR "edge native"). Title: (reliability OR reliable OR resilience OR resilient OR "fault tolerance" OR "fault tolerant" OR availability OR scalability OR scalable OR autoscaling OR "auto scaling" OR autoscaler OR elastic OR elasticity OR observability OR tracing OR monitoring OR "root cause" OR "anomaly detection" OR "chaos engineering" OR "fault injection" OR failure OR failures OR incident OR incidents OR "self healing" OR diagnosis OR debugging OR SLO OR "cold start" OR "performance variability" OR robustness OR dependability).

**S2.** Title and abstract: ("edge AI" OR "edge intelligence" OR "on-device inference" OR "on-device learning" OR "on-device analytics" OR TinyML OR "edge inference" OR "edge analytics" OR "edge machine learning"). Title: (reliability OR reliable OR resilience OR resilient OR "fault tolerance" OR "fault tolerant" OR robustness OR availability OR scalability OR scalable OR failure OR failures OR dependable OR dependability).

**S3.** Title: (agent OR agents OR agentic OR "multi-agent"). Title and abstract: ("large language model" OR "large language models" OR LLM OR LLMs). Title and abstract: (cloud OR microservice OR microservices OR AIOps OR incident OR "root cause" OR Kubernetes OR "site reliability").

**S4.** Title: (agentic OR "LLM agent" OR "LLM agents" OR "LLM-based agent" OR "LLM-based agents" OR "LLM-based multi-agent" OR "language agents"). Title: (reliability OR reliable OR robustness OR "fault tolerance" OR "fault tolerant" OR failure OR failures OR scalability OR scalable OR resilience OR resilient OR dependability). Title and abstract: (cloud OR edge OR microservice OR microservices OR serverless OR Kubernetes OR deployment OR production OR infrastructure OR distributed).

**S5.** Title and abstract: ("physical AI" OR "embodied AI" OR "cloud robotics" OR "fog robotics" OR "edge robotics" OR "robot offloading" OR "vision-language-action" OR "embodied intelligence"). Title and abstract: (edge OR fog OR offloading OR "cloud-edge" OR "edge-cloud"). Title and abstract: (reliability OR reliable OR "fault tolerance" OR scalability OR scalable OR resilience OR availability OR latency).

Records from other methods. An earlier IEEE Xplore, ACM Digital Library and Google Scholar search yielded 189 records, but its strings, dates and hit counts were not recorded. We treated it as an unreproducible earlier search, added its records under PRISMA 'other methods' and applied the same criteria.

Search validation. The database search retrieved 21 of 62 eligible earlier-search records (33.9%). Thirty-eight of 41 misses lack an S1 title term, showing the recall cost of title restriction. Adding the earlier records partly compensates; residual risk is addressed under threats to validity.

### 3.3 De-duplication and automated venue rule

The five blocks returned 7,112 records. Removing 117 cross-block duplicates and 184 duplicate DOIs or normalised titles left 6,811 records, a total of 301 removals. An automated venue rule then removed 2,341 records as E3 unless their DOI prefix identified a recognised publisher or their venue was a recognised peer-reviewed conference without publisher DOIs (Dataset S1). The rule verifies a publication channel, not venue quality; 4,470 database records remained.

Of 189 earlier-search records, 3 were duplicate rows, 31 overlapped the database search and 14 failed the venue rule, leaving 141. Thus 4,611 records entered screening, with each removal shown in Figure 1.

### 3.4 Eligibility criteria

A record was included when all of the following held:

- **I1.** It is a primary study that proposes, evaluates or empirically analyses a mechanism, tool, architecture or practice. Secondary studies (surveys, reviews and mapping studies) were logged as SR and used only for positioning.
- **I2.** It addresses reliability, resilience, fault tolerance, availability, scalability or elasticity, or the observability and diagnosis that support them, for cloud-native, edge-native or edge–cloud systems, or for AI workloads and agents running on such systems.
- **I3.** It reports an evaluation: a simulation, a testbed or benchmark experiment, an empirical study of production data, or a production deployment.
- **I4.** It is a peer-reviewed journal article, conference paper or book chapter, written in English and published from 2017 to 2026.

A record was excluded when any of the following held:

- **E1.** It is off-topic, that is, outside I2. During screening we recorded explicit rules for recurring borderline topics, all coded E1: hardware-level fault tolerance of neural-network accelerators; accuracy or privacy of federated learning without a system reliability or scalability concern; management of wireless, satellite, UAV or telecommunication networks; agent capability benchmarks and business, software-engineering or graphical-interface uses of agents without a reliability or scalability concern; and work that addresses only security or only energy.
- **E2.** It is a position, vision, demonstration or tutorial paper without an evaluation.
- **E3.** It is not peer-reviewed, its peer review cannot be verified (venue rule), or it was retracted.
- **E4.** It duplicates, or is a shorter version of, an included study.
- **E5.** It is not written in English.
- **E6.** It was published before 2017 (applied to the earlier-search records; the database blocks were restricted by year).
- **NR.** Its abstract could not be retrieved from the index or from the publisher's landing page, so it could not be assessed.

### 3.5 Five-point relevance scale

Each record that met I1 to I4 at the abstract stage was assigned a relevance level (Supplementary Table S7). Levels 3 to 5 were included; records judged at level 1 or 2 were excluded and coded E1. The level reflects how central the RQ topics are to the study's main contribution, not the study's quality, which was appraised separately. Of the 1,794 included studies, 621 were rated 5, 914 were rated 4 and 259 were rated 3. Relevance was also used, together with evidence level and the abstract-based reporting appraisal, to select the synthesis core set.

**Supplementary Table S7.** Five-point relevance scale and how it was applied.

### 3.6 Screening procedure and use of AI assistance

Screening had two stages. At the title stage, a record was excluded only when it was clearly outside I1 to I4; in case of doubt, it went on to the abstract stage. At the abstract stage, the full criteria and the relevance scale were applied, and included studies were coded (next subsection). For records without an abstract in the index, the abstract was sought from Semantic Scholar and from the publisher's landing page; 36 records whose abstract could not be obtained were recorded as reports not retrieved (NR).

The single author wrote the criteria, taxonomy and coding rules; an LLM (Claude, Anthropic) applied them to titles and abstracts and recorded decisions and exclusion codes. A keyword script proposed category, evidence and relevance codes, and LLM-assisted screening corrected 567 of 1,794 included records. No inter-rater reliability statistic is reported. Released decisions support audit but do not replace dual independent screening.

### 3.7 Classification

**Taxonomy.** The taxonomy was developed iteratively. An initial set of categories was derived from the research questions and from the related secondary studies (Supplementary Table S5); it was refined during abstract screening until every included study could be assigned exactly one primary category, the one that matches its main contribution. The final taxonomy has 13 categories (Table 3), grouped by research question: RQ1 covers C1, C4 and C5; RQ2 covers C2 and C3; RQ3 covers C6 to C9; and RQ4 covers C10 to C13.

**Evidence level.** Each included study was assigned the level of the strongest evaluation reported in its abstract (Supplementary Table S8), or EL? when the abstract did not allow a decision. A trace-driven simulation counts as EL1; a real system that replays production traces counts as EL3.

**Supplementary Table S8.** Evidence levels used to separate simulation, prototype and production evidence.

**Deployment scope.** To compare cloud and edge settings, each included study was also flagged as edge or edge–cloud when its title or abstract mentions edge, fog, IoT, devices, vehicles, drones, robots, multi-access edge computing or embedded and microcontroller hardware. This keyword-derived facet is approximate and is used only for descriptive comparisons.

### 3.8 Synthesis core set and abstract-based reporting appraisal

Each category was synthesised from a core set selected by explicit rules: relevance 5 (or 4 for sparse C4, C5 and C12), determinable evidence, quantitative abstract results except for empirical C5 studies, and preference for stronger evidence, citation rate and mechanism diversity. Earlier studies analysed by the author were retained when eligible. The result was 6–10 studies per category and 93 total (Dataset S1).

Six reporting items adapted from Dybå & Dingsøyr (2008) and Kitchenham & Charters (2007) were scored 1, 0.5 or 0 from indexed abstracts (Supplementary Table S9). Scores exist for 87 core studies; six earlier-search summaries were insufficient and are context only. No completed full-text reappraisal is claimed.

**Supplementary Table S9. Abstract-based reporting-appraisal items (each scored 1, 0.5 or 0).**

Findings require one study scoring at least 4.5 or two scoring at least 3; lower or unscored studies cannot support findings alone. Each quoted result carries its evidence level. Heterogeneous results are not pooled or ranked, and field-wide counts use all 1,794 studies.

## 4 RESULTS: THE SYSTEMATIC MAP

### 4.1 Search and selection

Figure 1 shows the selection process. Of the 4,611 records screened, 2,222 were excluded at the title stage, 36 could not be assessed because no abstract was available, and 559 were excluded at the abstract stage. The review includes 1,794 primary studies (1,753 from the database search and 41 from the earlier search) and logged 95 secondary studies. Off-topic records (E1) dominated the exclusions. Their share differed strongly between blocks: 70.6% of the screened S1 records were included, against 25.7% for S3, 23.8% for S5, 11.1% for S2 and 8.1% for S4. The AI-related terms in S2 to S5 also match robotics, multi-agent control, wireless networking and machine-learning papers that do not concern system reliability or scalability, which the E1 rules excluded.

**Figure 1.** PRISMA 2020 flow diagram of the search and selection. Counts are taken from Dataset S1; every removal step is shown separately, so the counts at consecutive stages add up.

### 4.2 Publication trend

Supplementary Figure S1 shows the included studies by year and research-question group. Output grew from 11 studies in 2017 to 417 in 2025, and 478 studies were indexed for 2026 up to the search date; 65.4% of all included studies were published from 2024 onwards. The growth is steepest for RQ4, because the categories on agentic AI (C11 and C12) consist entirely of studies published from 2024, when LLM-based agents for operations and their reliability became research topics.

**Supplementary Figure S1.** Included primary studies per publication year, by research-question group. The 2026 bar covers studies indexed up to the search date (11 October 2026).

### 4.3 Taxonomy and evidence levels

Table 3 and Figure 2 show the taxonomy and evidence. The largest categories are autoscaling and elasticity (C6, 408 studies), anomaly detection and root-cause analysis (C3, 351) and fault tolerance, availability and recovery (C1, 235). Evidence was determinable for 1,003 studies (55.9%): 66.2% prototypes or testbeds, 18.0% simulations, 11.2% production data and 4.6% production deployments. Thus 15.8% report production evidence.

The share of production evidence differs between categories in ways that follow from what each category studies. It is highest for empirical failure studies (C5, 87.5%), which analyse incidents, bugs and practitioner experience by design, and for agentic AI in operations (C11, 39.3%), where several systems were built and deployed by cloud operators. It is lowest for performance evaluation and benchmarking (C9, 2.7%), physical AI (C13, 2.9%) and resilience testing (C4, 4.8%), where laboratory testbeds are the norm, and it is also low for fault tolerance and recovery (C1, 8.8%).

Among C1–C9's 1,493 studies, 290 (19.4%) concern edge or edge–cloud deployments. Of studies with determinable evidence, simulation accounts for 26.9% at the edge versus 14.0% in cloud-only settings, and production evidence for 10.3% versus 15.5%. Edge scope is most frequent in scheduling and placement (32.3%) and fault tolerance (31.1%), and least frequent in RCA (9.1%) and benchmarking (7.3%). Serverless dominates cold-start studies (94.9%) and half of benchmarking studies (50.9%).

**Table 3.** Taxonomy of the 1,794 included primary studies with their evidence levels.

EL1 simulation or model; EL2 prototype or testbed; EL3 production data; EL4 production deployment; EL? not determinable from the abstract. "EL3+EL4 (%)" is the share of production evidence among studies with a determinable level. "Edge (%)" is the keyword-derived share of studies addressing edge or edge–cloud settings (see Review Method). The study-level coding is in Dataset S1, sheet Included_studies.

**Figure 2.** Taxonomy categories by evidence level: (a) number of included studies, including those whose level is not determinable from the abstract; (b) evidence mix among studies with a determinable level, with the share of production evidence (EL3+EL4) at the right.

### 4.4 Core set and abstract-based reporting appraisal

Supplementary Table S10 summarises the 93-study core: 18 EL4, 23 EL3, 46 EL2, 3 EL1 and 3 context-only studies with undetermined levels. Among 87 scored studies, the mean abstract-based reporting score is 5.05 of 6 (median 5.0); 73 score at least 4.5, 13 score 3–4 and 1 scores below 3. Mean item scores range from 1.00 for problem statement to 0.52 for artifact or data availability. These are abstract-reporting scores, not full methodological-quality assessments.

**Supplementary Table S10. Synthesis core set: evidence levels and abstract-based reporting scores by category.**

Abstract-based QA scores range from 0 to 6 (Supplementary Table S9). "Context only" marks studies without a complete abstract-based score or scoring below 3; these studies are cited but do not support findings on their own. Study-level scores and blank author-only full-text-check fields are in Dataset S1, sheet Core_set_QA.

## 5 RQ1: RELIABILITY AND FAULT TOLERANCE

RQ1 covers 281 studies (15.7% of the corpus): 235 on fault tolerance, high availability and recovery (C1), 35 on resilience testing and chaos engineering (C4) and 11 empirical failure studies (C5). Production data or deployments support 12.0% of the RQ1 studies whose evidence level is determinable. Supplementary Table S1 compares the core studies; the subsections below synthesise them by mechanism family.

### 5.1 Containing failure propagation and overload

Two production systems contain cascading failure. RPCShield found more than 3,800 risks across over 6,000 Uber microservices; owners remediated more than 1,300, with about 5% false positives (EL4) (Chabbi et al., 2026). DAGOR coordinated load shedding in WeChat for five years while preserving success and fairness under overload (EL4) (Zhou et al., 2018b). Evidence therefore supports both pre-deployment error analysis and run-time overload control, although the studies do not compare them.

### 5.2 Automated recovery and stateful fault tolerance

MicroRAS recovered 94.7% of faulty Kubernetes services while reducing interference by at least 44.3% (EL2) (Li et al., 2020a). Predictive migration reduced QoS violations for stateful pods (EL2) (Tran et al., 2022). DoubleTrack reduced serverless-workflow latency by up to 45.8% and logging overhead by 46.5% (EL2) (Bian et al., 2024). All are testbed results; none demonstrates automated stateful recovery in production.

### 5.3 Reliability-aware placement in the cloud and at the edge

GA-Par improved on a greedy placement method by 42.34% and ran about four times faster on real-world datasets (EL3) (Wen et al., 2019). At the edge, shared image layers increased simulated reliability by 12.4% and reduced delay by 28.57% (EL1) (Shi et al., 2024). Edge studies rely more on simulation, and only 3 edge or edge–cloud studies across C1–C9 report production deployments.

### 5.4 Changes that erode resilience

Software changes are a frequent trigger of failures. He et al. (2024) analysed 256 real incidents from four microservice systems and found that 37.87% of the erroneous software changes made the system less able to tolerate later faults, a problem that ordinary post-change monitoring misses because the fault may arrive much later. Their ResilienceGuardian framework injects faults in a staging environment and trains lightweight classifiers on pairs of KPI segments, identifying such changes with an F1-score of 0.9 (EL3). In a case study with about 1,000 microservices at a financial services company, repaying architectural technical debt, such as missing communication standards and poorly managed dead-letter queues, reduced the total number of incidents by 84% and critical and high-priority incidents by about 90% (EL4) (de Toledo et al., 2021). Both studies point to the same lever: reliability depends on how changes and architectural debt are managed, not only on run-time mechanisms.

### 5.5 Resilience testing and chaos engineering

Resilience testing is mostly laboratory based (19 EL2 and 1 EL4 among determinable studies). CrashFuzz found four unknown and one known crash-recovery bug (EL2) (Gao et al., 2023); IntelliFT guided recovery testing within a time budget (EL2) (Long et al., 2020). UncertainFuzz found 56% more new failure paths and reached 86% SLO-violation coverage versus 62% for rule-based and 48% for random injection (EL2) (Xue et al., 2026). FaultWeave found 237 vulnerabilities during three months on 512 industrial microservices (EL4) (Zheng et al., 2026).

Two studies test resilience in settings other than data-centre microservices. Chen et al. (2026) combined Chaos Mesh, Gremlin and ChaosBlade with workload generation and ran 11,965 fault-injection scenarios on Kubernetes: cloud–edge deployments showed about 80% greater response stability than cloud deployments under network delay and partition, whereas cloud deployments were about 47% more resilient under bandwidth limitation (EL2). The best deployment option therefore depends on the dominant fault type, which supports the distinction between cloud-native and edge-native failure models in Table 1. For serverless applications, Zayed & Al-Said Ahmad (2024) injected delays into AWS Lambda functions during code execution and at run time and compared performance with a fault-free baseline (EL2); their short paper reports no figures in its abstract.

### 5.6 What empirical failure studies show

Most of the 11 empirical failure studies use production data. Evidence includes 22 replicated industrial faults (EL2) (Zhou et al., 2018a), a survey of 106 practitioners plus six interviews (EL3) (Waseem et al., 2021), production-incident delays (EL3) (Ghosh et al., 2022), 210 bugs in 36 Kubernetes operators (EL3) (Xu et al., 2024) and Alibaba critical-path variability (EL3) (Ezaz et al., 2024). They identify slow diagnosis, complex dependencies, faulty automation and architectural debt. None studies an edge setting.

Supplementary Table S1. Core studies for RQ1 (reliability and fault tolerance).

EL: evidence level (Supplementary Table S8). QA: abstract-based reporting score out of 6 (Supplementary Table S9). Results are quoted from each study and are not comparable across studies, which use different systems, workloads and metrics.

## 6 RQ2: OBSERVABILITY, ANOMALY DETECTION AND ROOT-CAUSE ANALYSIS

RQ2 covers 508 studies (28.3% of the corpus): 157 on observability (C2) and 351 on anomaly detection and root-cause analysis (C3), the second-largest category of the map. Production data or deployments support 16.0% of the RQ2 studies with a determinable evidence level. Diagnosis research concentrates on microservices: only 9.1% of the C3 studies concern edge or edge–cloud settings. Supplementary Table S2 compares the core studies.

### 6.1 Collecting telemetry at acceptable cost

Practitioners at ten companies reported widespread tracing but limited machine learning (EL3) (Li et al., 2021a). TraceWeaver reconstructed traces at about 90% accuracy without application changes (EL3) (Ashok et al., 2024). eBPF collection reduced CPU overhead by up to 210 times and memory by up to 159% (EL2) (Sharma & Nadig, 2024). Across 11 probe patterns, targets used under 1% CPU while holders reached 10 GiB, and sharing improved efficiency (EL2) (Tundo et al., 2024). Lindorm combined high-rate storage, queries and SQL analytics (EL2) (Shen et al., 2023); STEAM is context only (He et al., 2023).

The finding is that the cost of observability grows with system size and that the reviewed mechanisms reduce it by avoiding application instrumentation, through kernel-level collection or trace reconstruction, and by sharing collection infrastructure. Most of this evidence comes from testbeds, and the practitioner study suggests that industrial analysis practice lags behind the analysis techniques proposed in research.

### 6.2 Root-cause analysis: graphs, traces and multimodal learning

Root-cause analysis (RCA) has the strongest production evidence among the diagnosis topics. Three families of techniques can be distinguished.

Dependency graphs localised benchmark faults at 89% precision and 97% mean average precision with MicroRCA (EL2) (Li et al., 2020b). CloudRanger improved accuracy by 10% on IBM Bluemix incidents and simulation (EL3) (Wang et al., 2018). At Alibaba, MicroHECL reached a 68% top-3 hit ratio and cut localisation time from 30 to 5 minutes (EL4) (Liu et al., 2021).

TraceRCA exceeded prior unsupervised top-1 accuracy by 44.8% and was applied at a commercial bank (EL4) (Li et al., 2021b). Seer anticipated 91% of QoS violations and avoided 84% in clusters (EL2) (Gan et al., 2019).

Multimodal systems improved RCA: DiagFusion by 20.9%–368% for instances and 11.0%–169% for failure types (EL3) (Zhang et al., 2023); Nezha to 89.77% mean top-1 accuracy (EL2) (Yu et al., 2023); and DeepTraLog to 0.93 precision and 0.97 recall (EL2) (Zhang et al., 2022). At eBay, LatentScope improved top-1 recall by 9.7%–57.9%, with public code and data (EL4) (Xie et al., 2024). Trace-log localisation is context only (Zhou et al., 2019b).

RCA is deployed at several operators, and multimodal data improves localisation. Testbed accuracy near 90% should not be equated with the 68% production top-3 result because systems and metrics differ; heterogeneous measures prevent ranking.

Supplementary Table S2. Core studies for RQ2 (observability, anomaly detection and root-cause analysis).

EL: evidence level (Supplementary Table S8). QA: abstract-based reporting score out of 6 (Supplementary Table S9); "–" marks studies cited for context only. Results are not comparable across studies.

## 7 RQ3: SCALABILITY AND RESOURCE MANAGEMENT

RQ3 is the largest group, with 704 studies (39.2% of the corpus): 408 on autoscaling and elasticity (C6), 117 on cold starts and performance variability (C7), 124 on scheduling, placement and orchestration (C8) and 55 on performance evaluation and benchmarking (C9). Production data or deployments support 14.6% of the RQ3 studies with a determinable evidence level. Supplementary Table S3 compares the core studies.

### 7.1 Autoscaling: from threshold rules to models and learning

At Alibaba, AHPA raised CPU use by 10% and cut resource cost by more than 20% (EL4) (Zhou et al., 2023); proactive–reactive provisioning improved resource use by 10%–15% (EL4) (Xu et al., 2023). In testbeds, ATOM improved throughput by about 30%–37% (EL2) (Gias et al., 2019), CoScal cut response time by 19%–29% and raised transactions by 6%–10% (EL2) (Xu et al., 2022), and a random-forest controller reached 90% utilisation and improved P95 latency by up to 95% (EL2) (Al Qassem et al., 2023). HGraphScale cut mean response time by up to 80.16% on real traces (EL3) (Fang et al., 2026).

The trade-off between SLO compliance and cost is explicit in a burst-aware autoscaler: in trace-driven simulation, it processed 1.09 times as many requests and reduced SLO violations by a factor of 5.17, but at a higher cost than the baseline (EL1) (Abdullah et al., 2020). For serverless workflows, StepConf configures the memory of each function step while the workflow runs and saved up to 40.9% of the cost on AWS Lambda while meeting the SLOs (EL2) (Wen et al., 2022). An in-production characterisation of an open-source serverless platform motivates new scaling strategies (cited for context because its indexed abstract is incomplete) (Nasiri et al., 2026).

Two findings follow. Predictive and model-based autoscalers outperform reactive threshold rules, and two deployments at one large operator report resource savings of 10% to 20% in production. Learning-based autoscalers report larger gains, but only in testbeds and trace replays, and their gains depend on the baseline and workload chosen.

### 7.2 Cold starts and performance variability in serverless platforms

A Huawei trace of 85 billion requests and 11.9 million cold starts found delays up to 7 seconds in one region and 3 seconds in another (EL3) (Joosen et al., 2025). Across 49,500 AWS and Azure executions, overhead ranged from 300 ms to 24 s (EL2) (Manner et al., 2018). FaaSLight cut code-loading latency by up to 78.95% (28.78% mean) and total latency by up to 42.05% (19.21% mean) (EL2) (Liu et al., 2023). Incendio was 1.4 times faster and cut latency by 23% and 14.8% versus two systems (EL3) (Cai et al., 2024). FlashServe accelerated cold starts by up to 49 times versus object storage and 3.3 times versus ServerlessLLM, cut GPU idle cost by 32% and kept 95% of time-to-first-token below one second (EL3) (Chen, 2025).

Performance also varies over time. Months of measurements on Google Cloud Functions showed diurnal performance differences of up to 15% and three times more unexpected cold starts at the start of the week (EL2) (Schirmer et al., 2023). Further studies on run-to-run variance and on the stability of serverless performance tests are cited for context (Eismann et al., 2022; Wen et al., 2025).

The findings are that cold-start latency varies by an order of magnitude or more with language, memory, artifact size, trigger type and region, and that application-level and platform-level mitigations reduce it substantially in testbeds and trace replays. No study in C7 was coded as a production deployment (EL4), so these mitigations have not been shown to work in production. Variability over time also threatens the validity of serverless experiments themselves, which matters for interpreting the results above.

### 7.3 Scheduling, placement, service composition and configuration

Alibaba's C-Koordinator predicted interference above 90.3% accuracy and improved P50–P99 latency by 16.7%–36.1% (EL4) (Song et al., 2026). Erms reduced SLA-violation probability fivefold and resources 1.6-fold (EL2) (Luo et al., 2023). ComboFunc scaled up to 3.01 times faster and cut cost by up to 42.6% (EL3) (Wen et al., 2024). Paldia improved SLO compliance by up to 13.3%, cut tail latency by about 50% and saved up to 86% (EL3) (Bhasi et al., 2024).

Service composition and configuration appear in three core studies. orcon makes relationships between services first-class metadata in the Kubernetes API, based on lessons from running large smart-city platforms, and propagated lifecycle and configuration changes with an overhead of 0.44 seconds per service (EL4) (Sebrechts et al., 2021). SLAM derives the structure of a serverless application from distributed traces and chooses memory configurations so that more than 95% of requests met the SLO on AWS Lambda (EL2) (Safaryan et al., 2022), and StepConf, discussed above, configures each step of a function workflow at run time (Wen et al., 2022). The findings are that interference among co-located services is a first-order limit to scalability, which interference-aware co-location reduces in production, and that configuring compositions of functions per step or per function from tracing data meets SLOs at lower cost than uniform configuration.

### 7.4 Benchmarking scalability

Benchmarking has 2.7% production evidence. One method found that at most five repetitions of five minutes could test SLO compliance (EL2) (Henning & Hasselbring, 2022). Over 740 hours, five stream frameworks scaled roughly linearly with enough resources but required different capacity (EL2) (Henning & Hasselbring, 2023). A monolith outperformed microservices on one machine; vertical scaling cost less than horizontal scaling, whose performance eventually declined (EL2) (Blinowski et al., 2022). Serverless suites are context only (Shahrad et al., 2019; Copik et al., 2021).

The finding is that scalability is neither free nor unlimited: microservices pay an overhead at small scale, horizontal scaling has limits, and the choice of framework changes the cost of scaling. Benchmark methods that test SLO compliance per load and resource combination make such statements reproducible at modest cost.

Supplementary Table S3. Core studies for RQ3 (scalability and resource management).

EL: evidence level (Supplementary Table S8). QA: abstract-based reporting score out of 6 (Supplementary Table S9); "–" marks studies cited for context only. Results are not comparable across studies.

## 8 RQ4: EDGE AI, AGENTIC AI AND PHYSICAL AI

RQ4 covers 301 studies (16.8% of the corpus): 64 on edge AI and on-device analytics (C10), 88 on agentic AI for cloud operations (C11), 77 on the reliability and scalability of agentic AI systems (C12) and 72 on physical AI on the edge–cloud continuum (C13). It is the youngest part of the map: 82.7% of its studies were published from 2024 onwards. Production data or deployments support 21.3% of the RQ4 studies with a determinable evidence level, a share driven by agentic AI for operations. Supplementary Table S4 compares the core studies.

### 8.1 Edge AI and on-device analytics

Edge AI uses inference-specific redundancy and joint adaptation. Flake reduced maximum-to-median latency disparity by up to 2.5 times (EL2) (Liu et al., 2022). Multi-model placement reduced cost and QoS loss under node failure (EL2) (Wu et al., 2022). EdgeAdaptor stayed within an empirical competitive ratio of 1.4 (EL1) (Zhao et al., 2022), rolling-horizon optimisation scheduled industrial models and resources (EL2) (Li et al., 2025), and Explainability-as-a-Service reduced latency by 38% in three deployments (EL4) (Singh & Roy, 2026).

Field evidence is rare but instructive. A post-mortem of an edge AI wildlife-deterrence system reports a total energy collapse within 24 hours, caused by fog and condensation that defeated the solar and optical design and by "trigger storms" of non-target animals that kept an always-on GPU busy; an event-driven design with cascaded wake-up and thermal vision restored operation (EL4) (Hajder et al., 2026). This single study illustrates failure modes, such as energy depletion and environmental effects, that testbed studies of edge AI do not model, which is consistent with the failure model of edge-native systems in Table 1.

On-device analytics in the strict sense, with inference on the end device itself, is a small part of the category: 15 of the 64 C10 studies mention on-device, mobile, wearable or microcontroller-class execution in their title or abstract. The opposite direction, AI used to manage edge resources (AI-driven edge intelligence), is more common than the C10 count suggests. Among the 293 edge or edge–cloud studies on fault tolerance, autoscaling, cold starts, placement and physical AI (C1, C6 to C8 and C13), 83 (28.3%) use reinforcement learning, deep learning or other machine-learning techniques to make management decisions, for example the learned offloading policies discussed under physical AI below.

The finding for edge AI is that reliability is pursued through redundancy adapted to inference (coded computation, composition of several models with load balancing) and scalability through joint adaptation of models, configurations and resources. The evidence comes mainly from simulations and testbeds; the two production studies in the core set report a latency reduction and a field failure, not a validated fault-tolerance mechanism.

### 8.2 Agentic AI for cloud operations

Four agentic-operations systems report production use. KRCA reached 0.88 accuracy for service and 0.79 for failure type, at least 31% above baseline, and cut diagnosis time by 77.3% over six months (EL4) (Jiang et al., 2026). XiHe assisted over 3,000 incidents in 12 months, cutting root-cause time by 25.8% with 94.6% accuracy and 95.1% satisfaction (EL4) (Ren et al., 2026). OncallX responded in 21 seconds and triaged in 4 seconds (EL4) (Fu et al., 2025b). DualLane reported 96.5% offline accuracy, 4.2-second median plan execution and 7.1% agent-induced errors after over a year (EL4) (Wang et al., 2026a). RCAgent and a ReAct agent also used production incidents (EL3) (Wang et al., 2024c; Roy et al., 2024).

These studies also show how agents are made reliable enough to operate other systems. The first lever is grounding in tools and live observations: in a closed-loop testbed on a live 5G core deployed on Kubernetes, giving LLM agents tool access raised their average remediation success from 10.9% to 61.1% across 900 experiments, while model size alone did not guarantee success (EL2) (de Castro et al., 2026). The second is structure that constrains reasoning, such as causal skeletons (Jiang et al., 2026) and verifiable causal graphs (Ren et al., 2026). The third is explicit safety properties. STRATUS organises detection, diagnosis and mitigation agents as a state machine and enforces a transactional no-regression property, so that a failed mitigation can be rolled back to a safe state; on the AIOpsLab and ITBench benchmarks, its mitigation success rate was at least 1.5 times that of earlier SRE agents (EL2) (Chen et al., 2025). The fourth is bounding cost and error for frequent cases with pre-validated execution templates (Wang et al., 2026a).

The findings are that LLM agents for operations are in production use at several operators, with large reductions in diagnosis and response time, and that their reliability rests on grounding in tools and live observations, on structural constraints and on explicit safety properties. A caution applies: every production report comes from the operator that built the agent, and no independent replication or common benchmark connects the production results.

### 8.3 Reliability and scalability of agentic AI applications

Agent applications fail in framework-specific ways. Studies found 1,026 bugs in 1,577 pull requests (EL3) (Xue et al., 2025), 409 fixed bugs across five multi-agent frameworks with tests in only 47% of model-integration fixes (EL3) (Zhang et al., 2026), and 547 confirmed coding-agent safety failures among 16,586 issues, including 326 high or critical and over 65% during bug fixing or setup (EL3) (Al Hasan & Biswas, 2026). Across 1,980 tool tests, a mid-sized model reached 96.6% success at 7.3 seconds per task on commodity hardware (EL2) (Huang et al., 2026).

Reliability is then ensured with mechanisms that mirror cloud-native practice. AgentTrace defines a seven-class failure taxonomy, normalises heterogeneous agent events into traces and repairs failures at run time without additional LLM calls; on 300 interactive tasks and a subset of WebArena it raised task success from 52.4% to 68.9% and more than halved execution-critical failures, with less than 5% latency overhead (EL2) (Kothari et al., 2026). ResMAS showed that the resilience of LLM multi-agent systems to perturbations such as agent failures depends on their communication topology and prompts, and improved it by generating resilient topologies and topology-aware prompts (EL2) (Zhou et al., 2026). The transactional no-regression property of STRATUS (Chen et al., 2025), discussed above, plays the role of a rollback mechanism.

Scalability is limited by the multi-turn structure of agent workloads. A system-level characterisation of AI agents found that accuracy improves with more test-time compute but with diminishing returns, widening latency variance and rising data-centre power demand (EL2) (Kim et al., 2026). Maestro predicts the output length and memory use of each agent stage and uses the prediction for node-level co-location of models, cluster-level routing that avoids cold starts and workflow-aware prioritisation; it reduced the high-bandwidth memory reserved for key-value caches by 67.2% and improved SLO attainment under high contention by 23.6 percentage points over earliest-deadline-first scheduling (EL2) (Wang et al., 2026b). An analysis of agent workloads on Kubernetes identified GPU memory, orchestration overhead, storage I/O and inter-agent communication as bottlenecks and reported 2 to 5 times higher throughput and 33 times faster cold starts after optimisation in production deployments (EL4) (Mittal & Tadi, 2025).

Taken together, the evidence supports the following practices for ensuring the reliability and scalability of agentic AI applications, each with at least the support required by our synthesis rule: (i) treat the agent framework as a dependency that fails, and test its model-integration and orchestration layers (Xue et al., 2025;

Zhang et al., 2026); (ii) trace agent executions and repair known failure classes automatically (Kothari et al., 2026); (iii) give agents tool access and verify their actions in a closed loop against the live system (de Castro et al., 2026); (iv) enforce safety properties that allow rollback of agent actions (Chen et al., 2025); and (v) schedule agent workflows with knowledge of their stage-level resource demands (Wang et al., 2026b). The evidence for these practices comes mainly from repositories and testbeds; production evidence for agent applications, as distinct from agents that operate clouds, is limited to one study (Mittal & Tadi, 2025).

### 8.4 Physical AI on the edge–cloud continuum

Physical AI scales through edge and cloud offloading, with reliability secondary in 52 of 72 C13 studies and no level-5 relevance ratings. Network-aware offloading improved vision tasks by 1.3–2.3 times (EL2) (Chinchali et al., 2021). RecSLAM cut latency by up to 39.31% (EL2) (Huang et al., 2022). FogROS2 cut SLAM latency by 50%, grasp planning from 14 to 1.2 seconds, motion planning 45-fold, network use up to 3.8-fold and compressed-image round trips by 97% (EL2) (Ichnowski et al., 2023). Kubernetes edge robotics was fault tested (EL2) (Zhang et al., 2025a), algorithm allocation used production data (EL3) (Alirezazadeh et al., 2022), and CloudGripper provides 32 robot cells (EL2) (Zahid & Pokorny, 2024).

The finding is that offloading to edge or fog resources is the main lever for scaling physical AI, and that learned, network-aware offloading and container orchestration of robot software are the main means of keeping it dependable. The evidence is almost entirely from testbeds (2.9% production evidence), and no core study evaluates physical AI reliability in a production deployment.

Supplementary Table S4. Core studies for RQ4 (edge AI, agentic AI and physical AI).

EL: evidence level (Supplementary Table S8). QA: abstract-based reporting score out of 6 (Supplementary Table S9). Results are not comparable across studies.

## 9 DISCUSSION

### 9.1 How the evidence supports the taxonomy

The taxonomy is supported by complete assignment and systematic evidence differences (Table 3), from 2.7% production evidence in C9 to 87.5% in C5. Agentic operations (C11, 39.3%) and agent applications (C12, 22.2%) also differ. Cloud–edge fault response depends on fault type (Chen et al., 2026), field failures include energy and environment (Hajder et al., 2026), and edge evidence is weaker overall. Thin C4 and C5 categories may change as they grow.

### 9.2 How the evidence supports the conclusions and research directions

Supplementary Table S11 connects each main conclusion to the core studies that support it, to the map-level evidence and to the research direction it motivates. Every conclusion satisfies the synthesis rule stated in the method, and every research direction responds to a gap that is visible in the map rather than to an opinion about where the field should go.

Supplementary Table S11. Main conclusions, their supporting evidence and the research directions they motivate.

All conclusions are limited to the 1,794 studies documented in Dataset S1. Absence of evidence in this corpus is not evidence that a mechanism is ineffective or untested elsewhere.

Two tensions recur. Production deployments provide strong evidence but rarely release data or tools, while benchmarks may not represent production; LatentScope combines deployment with public code and data (Xie et al., 2024). Automation also creates failures (Xu et al., 2024; Zhang et al., 2026), so no-regression, verification and rollback must accompany recovery and remediation (Li et al., 2020a; Chen et al., 2025; de Castro et al., 2026).

### 9.3 Implications for practitioners

Within the limits of the corpus, the evidence supports the following practices: analyse error propagation and control overload across services before relying on per-service retries (Zhou et al., 2018b; Chabbi et al., 2026); test whether changes reduce fault resilience before release (He et al., 2024); combine metrics, logs and traces for diagnosis and do not expect testbed accuracy in production (Liu et al., 2021; Zhang et al., 2023); prefer predictive or model-based autoscaling and make the cost–SLO trade-off explicit (Abdullah et al., 2020; Zhou et al., 2023); measure cold starts per runtime, memory setting and region (Manner et al., 2018; Joosen et al., 2025); for edge and cloud–edge deployments, test network partitions and bandwidth limits separately (Chen et al., 2026); and, for LLM agents that operate or form part of a system, provide tools and closed-loop verification, trace and repair agent failures, and keep agent actions reversible (Chen et al., 2025; Kothari et al., 2026; de Castro et al., 2026).

### 9.4 Scope of the conclusions

All conclusions in this paper concern the 1,794 primary studies documented in Dataset S1, and the synthesis rests on the 93 core studies appraised in Supplementary Table S10. Statements about the field refer to counts in the map, which depend on what the logged search retrieved and on evidence levels determined from abstracts. Where the map shows little evidence, for example production evidence for edge-native reliability or physical AI, we report a gap in the documented literature, not a judgement that the mechanisms do not work.

## 10 THREATS TO VALIDITY

We discuss threats to validity following the categories proposed for secondary studies in software engineering by Ampatzoglou et al. (2019).

Study identification. A single index and title-restricted S1 may miss studies. OpenAlex retrieved 21 of 62 eligible earlier-search studies (33.9%); 38 of 41 misses lacked a required title term. Earlier records recovered those 41, but other work may be absent. The venue rule removed 2,341 unverifiable records and may exclude valid publishers while admitting uneven outlets. Snowballing (Wohlin, 2014) could improve recall. Thirty-six records lacked abstracts, and 2026 coverage ends at the search date.

Selection and extraction. One author used LLM assistance rather than dual independent screening; no completed human consistency estimate exists. Released rules and decisions permit audit but do not remove systematic-error risk. Abstract coding left evidence undetermined for 44.1% of studies, and keyword-derived facets are approximate.

Reporting appraisal. Six items score reporting visible in abstracts or retained summaries, not full methods, venue quality, sample size or statistical rigour. Six studies were unscored and context only. The synthesis rule limits unsupported claims but cannot replace full-text appraisal.

Synthesis. Core-set criteria favour relevance, determinable evidence and citation rate, which may over-represent large operators and some newer studies. Builder-reported production results are rarely replicated, positive-result bias is likely, and heterogeneous systems and metrics prevent pooling or ranking.

Conclusion validity and generalisability. The conclusions are limited to the documented corpus (Supplementary Table S11). The topics of RQ4, especially agentic AI, are changing quickly, and the studies of 2025 and 2026 dominate those categories, so these conclusions will date faster than those for RQ1 to RQ3.

## 11 CONCLUSIONS

This systematic mapping study documents 1,794 primary studies on the reliability and scalability of cloud-native, edge-native and AI-enabled systems, published from 2017 to 2026, and synthesises a 93-study core set appraised from indexed abstracts. Within the limits of this corpus, the answers to the research questions are as follows.

For RQ1, failure propagation and overload in large microservice systems can be contained in production by analysing error propagation and by coordinated load shedding; automated recovery, stateful fault tolerance and guided resilience testing improve on platform defaults in testbeds, with one industrial deployment of resilience testing. For RQ2, observability costs are reduced by instrumentation-free and kernel-level collection, and root-cause localisation has been deployed by several operators; multimodal methods localise faults better than single-modality ones, but testbed accuracy overstates production accuracy. For RQ3, predictive and model-based autoscaling outperforms threshold rules and has saved 10% to 20% of resources in production at one operator; cold-start mitigations and interference-aware placement are effective in testbeds and trace replays, and scalability benchmarks show that scaling is neither free nor unlimited. For RQ4, edge AI relies on redundancy adapted to inference and on joint adaptation of models and resources, with little production evidence; LLM agents for operations are in production at several operators and are made dependable through tool grounding, structural constraints and safety properties; agentic applications fail in agent-specific ways and are made more reliable by tracing and repair, closed-loop verification, reversible actions and workload-aware scheduling; and physical AI scales through edge and fog offloading, with reliability rarely studied as the main question.

Across all research questions, only 15.8% of the studies with a determinable evidence level report production evidence, and edge studies rely on simulation about twice as often as cloud-only studies. The research directions in Supplementary Table S11 follow from these gaps: shared and realistic benchmarks for resilience and for agentic operations, empirical failure studies of edge deployments, production evaluations of cold-start and autoscaling mechanisms, and dependability engineering for agent frameworks and physical AI. Dataset S1 allows other researchers to verify, extend and update this map.

## 12 DATA AVAILABILITY

Dataset S1 (Excel workbook) contains the search log with the exact query filters and hit counts; all 7,184 retrieved records with their de-duplication and venue-rule outcomes; the screening stage, decision, exclusion reason, relevance level, category and evidence level of each of the 4,611 screened records; the 1,794 included primary studies with bibliographic data and DOIs; the 95 secondary studies; the synthesis core set with its abstract-based reporting scores; the codebook and the venue rule; and formulas that reproduce the counts of Figure 1 and Table 3. Figures 1 and 2 and Supplementary Figure S1 are generated from these data.

## 13 USE OF AI TOOLS

The author used Claude (Anthropic), a large language model, to assist with applying the eligibility criteria and relevance scale to titles and abstracts, coding the included studies, pre-scoring the reporting-appraisal items from abstracts, writing the scripts that produced Dataset S1 and the figures, and editing the manuscript. The criteria, rules and all decisions are documented in Dataset S1. No inter-rater reliability statistic or completed full-text quality verification is claimed. The author reviewed the resulting manuscript and takes full responsibility for its content.

## References

[1] Abdullah M, Iqbal W, Berral JL, Polo J, Carrera D. 2020. Burst-Aware Predictive Autoscaling for Containerized Microservices. IEEE Transactions on Services Computing 15(3):1448–1460. DOI 10.1109/tsc.2020.2995937

[2] Afrin M, Jin J, Rahman A, Rahman A, Wan J, Hossain E. 2021. Resource Allocation and Service Provisioning in Multi-Agent Cloud Robotics: A Comprehensive Survey. IEEE Communications Surveys & Tutorials 23(2):842–870. DOI 10.1109/comst.2021.3061435

[3] Al Hasan A, Biswas S. 2026. What Breaks When LLMs Code? Characterizing Operational Safety Failures of Agentic Code Assistants. In: Proceedings of the IEEE/ACM International Conference on Automated Software Engineering (ASE 2026). New York: ACM, 790–802. DOI 10.1145/3832783.3834393

[4] Al Qassem LMMS, Stouraitis T, Damiani E, Elfadel IM. 2023. Proactive Random-Forest Autoscaler for Microservice Resource Allocation. IEEE Access 11:2570–2585. DOI 10.1109/access.2023.3234021

[5] Alirezazadeh S, Correia A, Alexandre LA. 2022. Optimal algorithm allocation for robotic network cloud systems. Robotics and Autonomous Systems 154:104144. DOI 10.1016/j.robot.2022.104144

[6] Ampatzoglou A, Bibi S, Avgeriou P, Verbeek M, Chatzigeorgiou A. 2019. Identifying, categorizing and mitigating threats to validity in software engineering secondary studies. Information and Software Technology 106:201–230. DOI 10.1016/j.infsof.2018.10.006

[7] Ashok S, Harsh V, Godfrey PB, Mittal R, Parthasarathy S, Shwartz L. 2024. TraceWeaver: Distributed Request Tracing for Microservices Without Application Modification. In: Proceedings of the ACM SIGCOMM Conference (SIGCOMM 2024). New York: ACM, 828–842. DOI 10.1145/3651890.3672254

[8] Avizienis A, Laprie J-C, Randell B, Landwehr C. 2004. Basic concepts and taxonomy of dependable and secure computing. IEEE Transactions on Dependable and Secure Computing 1(1):11–33. DOI 10.1109/TDSC.2004.2

[9] Bhasi VM, Sharma A, Mohanty S, Kandemir M, Das CR. 2024. Paldia: Enabling SLO-Compliant and Cost-Effective Serverless Computing on Heterogeneous Hardware. In: 2024 IEEE International Parallel and Distributed Processing Symposium (IPDPS). Piscataway: IEEE, 100–113. DOI 10.1109/ipdps57955.2024.00018

[10] Bian S, Wang W, Yang Y, Zhao L. 2024. DoubleTrack: Fault-Tolerant Stateful Serverless Computing with Asynchronous Shared Logs. In: Proceedings of the IEEE International Conference on High Performance Computing and Communications (HPCC 2024). Piscataway: IEEE, 806–813. DOI 10.1109/hpcc64274.2024.00111

[11] Blinowski G, Ojdowska A, Przybyłek A. 2022. Monolithic vs. Microservice Architecture: A Performance and Scalability Evaluation. IEEE Access 10:20357–20374. DOI 10.1109/access.2022.3152803

[12] Burns B, Grant B, Oppenheimer D, Brewer E, Wilkes J. 2016. Borg, Omega, and Kubernetes. Communications of the ACM 59(5):50–57. DOI 10.1145/2890784

[13] Cai XT, Sang Q, Hu C, Gong Y, Suo K, Zhou X, Cheng D. 2024. Incendio: Priority-Based Scheduling for Alleviating Cold Start in Serverless Computing. IEEE Transactions on Computers 73(7):1780–1794. DOI 10.1109/tc.2024.3386063

[14] Chabbi M, Mahajan S, Beschastnikh I, Just R, Wang Y, Xu Y, Pinto EM, Saha S, Sridharan M, Jha A, Sheshadri SK, Bansal M. 2026. RPCShield: Defending Microservices Against Cascading Failures. In: Proceedings of the ACM SIGOPS Symposium on Operating Systems Principles (SOSP 2026). New York: ACM, 802–826. DOI 10.1145/3830418.3843878

[15] Chen B. 2025. FlashServe: Cost-Efficient Serverless Inference Scheduling for Large Language Models via Tiered Memory Management and Predictive Autoscaling. In: Proceedings of the International Conference on Computer Science and Management Technology (ICCSMT 2025). New York: ACM, 1291–1298. DOI 10.1145/3795154.3795363

[16] Chen Z, Goudarzi M, Toosi AN. 2026. An Empirical Evaluation of Kubernetes Resilience in Cloud–Edge Deployments Using Failure Injection Techniques. In: Proceedings of the Australasian Computer Science Week (ACSW 2026). New York: ACM, 22–31. DOI 10.1145/3793811.3793823

[17] Chen Y, Pan J, Clark J, Su Y, Zheutlin N, Bhavya B, Arora R, Deng Y, Jha S, Xu T. 2025. STRATUS: A Multi-agent System for Autonomous Reliability Engineering of Modern Clouds. In: Advances in Neural Information Processing Systems 38 (NeurIPS 2025), 55835–55881. DOI 10.52202/085713-1671

[18] Chinchali S, Sharma A, Harrison J, Elhafsi A, Kang DD, Pergament E, Cidon E, Katti SR, Pavone M. 2021. Network Offloading Policies for Cloud Robotics: A Learning-Based Approach. Autonomous Robots 45(7):997–1012. DOI 10.1007/s10514-021-09987-4

[19] Cloud Native Computing Foundation. 2018. CNCF cloud native definition v1.0. Available at https://github.com/cncf/toc/blob/main/DEFINITION.md.

[20] Copik M, Kwasniewski G, Besta M, Podstawski M, Hoefler T. 2021. SeBS: A Serverless Benchmark Suite for Function-as-a-Service Computing. In: Proceedings of the 22nd International Middleware Conference (Middleware 2021). New York: ACM. DOI 10.1145/3464298.3476133

[21] de Castro AG, Vandikas K, Ferlin S, Chiesa M, Rothenberg CE. 2026. OperAID: Benchmarking LLM Agents for Autonomous Kubernetes Fault Remediation. In: Proceedings of the IEEE International Conference on Network Softwarization (NetSoft 2026). Piscataway: IEEE, 625–631. DOI 10.1109/netsoft70012.2026.11603505

[22] de Toledo SS, Martini A, Sjøberg DIK, Przybyszewska A, Frandsen JS. 2021. Reducing Incidents in Microservices by Repaying Architectural Technical Debt. In: Proceedings of the Euromicro Conference on Software Engineering and Advanced Applications (SEAA 2021). Piscataway: IEEE, 196–205. DOI 10.1109/seaa53835.2021.00033

[23] Duan J, Yu S, Tan HL, Zhu H, Tan C. 2022. A survey of embodied AI: from simulators to research tasks. IEEE Transactions on Emerging Topics in Computational Intelligence 6(2):230–244. DOI 10.1109/TETCI.2022.3141105

[24] Dybå T, Dingsøyr T. 2008. Empirical studies of agile software development: a systematic review. Information and Software Technology 50(9–10):833–859. DOI 10.1016/j.infsof.2008.01.006

[25] Eismann S, Costa DE, Liao L, Bezemer C-P, Shang W, van Hoorn A, Kounev S. 2022. A case study on the stability of performance tests for serverless applications. Journal of Systems and Software 189:111294.

[26] Ezaz A, Khodabandeh G, Ezzati-Jivan N. 2024. Analyzing Performance Variability in Alibaba's Microservice Architecture: A Critical-Path-Based Perspective. In: Proceedings of the ACM/SPEC International Conference on Performance Engineering (ICPE 2024). New York: ACM, 82–86. DOI 10.1145/3629527.3651845

[27] Fang Z, Ma H, Chen G, Buyya R. 2026. HGraphScale: Hierarchical Graph Learning for Autoscaling Microservice Applications in Container-Based Cloud Computing. IEEE Transactions on Services Computing 19(1):410–422. DOI 10.1109/tsc.2026.3651267

[28] Fu N, Cheng G, Teng Y, Dai G, Yu S, Chen Z. 2025a. Intelligent Root Cause Localization in MicroService Systems: A Survey and New Perspectives. ACM Computing Surveys 57(12):1–37. DOI 10.1145/3736755

[29] Fu R, Zhang Y, Che Z, Wu X, Zhong Z, Ren Z, Zhang S, Wang F, Sun Y, Liu X, Liu K, Zhang Y. 2025b. LLM-Powered Multi-Agent Collaboration for Intelligent Industrial On-Call Automation. In: Proceedings of the IEEE/ACM International Conference on Automated Software Engineering (ASE 2025). Piscataway: IEEE, 2222–2234. DOI 10.1109/ase63991.2025.00184

[30] Gan Y, Zhang Y, Hu K, Cheng D, He Y, Pancholi M, Delimitrou C. 2019. Seer: Leveraging Big Data to Navigate the Complexity of Performance Debugging in Cloud Microservices. In: Proceedings of the Twenty-Fourth International Conference on Architectural Support for Programming Languages and Operating Systems (ASPLOS 2019). New York: ACM. DOI 10.1145/3297858.3304004

[31] Gao Y, Dou W, Wang D, Feng W, Wei J, Zhong H, Huang T. 2023. Coverage Guided Fault Injection for Cloud Systems. In: Proceedings of the 45th IEEE/ACM International Conference on Software Engineering (ICSE 2023). Piscataway: IEEE. DOI 10.1109/icse48619.2023.00186

[32] Ghosh S, Shetty M, Bansal C, Nath S. 2022. How to fight production incidents?: an empirical study on a large-scale cloud service. In: Proceedings of the 13th Symposium on Cloud Computing (SoCC 2022). New York: ACM. DOI 10.1145/3542929.3563482

[33] Gias AU, Casale G, Woodside M. 2019. ATOM: Model-Driven Autoscaling for Microservices. In: Proceedings of the IEEE International Conference on Distributed Computing Systems (ICDCS 2019). Piscataway: IEEE, 1994–2004. DOI 10.1109/icdcs.2019.00197

[34] Golec M, Walia GK, Kumar M, Cuadrado F, Gill SS, Uhlig S. 2024. Cold Start Latency in Serverless Computing: A Systematic Review, Taxonomy, and Future Directions. ACM Computing Surveys 57(3):1–36. DOI 10.1145/3700875

[35] Gomes F, Rêgo PAL, Trinta FAM. 2025. A systematic mapping study on observability of microservices-based applications: fundamentals, classifications, and challenges. Computing 107(9). DOI 10.1007/s00607-025-01540-w

[36] Hajder M, Hajder P, Liput M, Kolbusz J, Rogólski R, Kiszkowiak Ł, Hajder L. 2026. Smart City Myths in the Wild: A Post-Mortem of an Edge AI Energy Failure in Wildlife Deterrence. In: Proceedings of the IEEE International Conference on Pervasive Computing and Communications Workshops and other Affiliated Events (PerCom Workshops 2026). Piscataway: IEEE, 1–6. DOI 10.1109/percomworkshops68308.2026.11585319

[37] He S, Feng B, Li L, Zhang X, Kang Y, Lin Q, Rajmohan S, Zhang D. 2023. STEAM: Observability-Preserving Trace Sampling. In: Proceedings of the 31st ACM Joint European Software Engineering Conference and Symposium on the Foundations of Software Engineering (ESEC/FSE 2023). New York: ACM. DOI 10.1145/3611643.3613881

[38] He G, Nie X, Tang R, Wang K, Yu Z, Wen X, Yin K, Pei D. 2024. Guardian of the Resiliency: Detecting Erroneous Software Changes Before They Make Your Microservice System Less Fault-Resilient. In: Proceedings of the IEEE/ACM International Symposium on Quality of Service (IWQoS 2024). Piscataway: IEEE, 1–10. DOI 10.1109/iwqos61813.2024.10682951

[39] Henning S, Hasselbring W. 2022. A configurable method for benchmarking scalability of cloud-native applications. Empirical Software Engineering 27(6). DOI 10.1007/s10664-022-10162-1

[40] Henning S, Hasselbring W. 2023. Benchmarking scalability of stream processing frameworks deployed as microservices in the cloud. Journal of Systems and Software 208:111879. DOI 10.1016/j.jss.2023.111879

[41] Herbst NR, Kounev S, Reussner R. 2013. Elasticity in cloud computing: what it is, and what it is not. In: 10th International Conference on Autonomic Computing (ICAC 13). San Jose: USENIX, 23–27.

[42] Huang D, Malwe G, Wang Z. 2026. When Agents Fail to Act: A Diagnostic Framework for Tool Invocation Reliability in Multi-Agent LLM Systems. In: Proceedings of the International Conference on Artificial Intelligence and Big Data (ICAIBD 2026). Piscataway: IEEE, 492–497. DOI 10.1109/icaibd69640.2026.11637248

[43] Huang P, Zeng L, Chen X, Huang L, Zhou Z, Yu S. 2022. Edge Robotics: Edge-Computing-Accelerated Multirobot Simultaneous Localization and Mapping. IEEE Internet of Things Journal 9(15):14087–14102. DOI 10.1109/jiot.2022.3146461

[44] Ichnowski J, Chen K, Dharmarajan K, Adebola S, Danielczuk M, Mayoral-Vilches V, Jha N, Zhan H, Llontop E, Xu DC, Buscaron C, Kubiatowicz JD, Stoica I, Gonzalez J, Goldberg K. 2023. FogROS2: An Adaptive Platform for Cloud and Fog Robotics Using ROS 2. In: 2023 IEEE International Conference on Robotics and Automation (ICRA). Piscataway: IEEE, 5493–5500. DOI 10.1109/icra48891.2023.10161307

[45] Jeong B, Jeong Y, Jeong Y. 2025. Autoscaling techniques in cloud-native computing: A comprehensive survey. Computer Science Review 58:100791. DOI 10.1016/j.cosrev.2025.100791

[46] Jiang J, Feng J, Luo Y, Zhang Q, Su YW, Gu W, Zhang S, Cui T, Wu Y, Huang J, Qi N, Pei D. 2026. KRCA: An Efficient Root Cause Analysis System in Hyper-scale Microservice Systems via Agentic AI. In: Proceedings of the IEEE/ACM International Conference on Automated Software Engineering (ASE 2026). New York: ACM, 3804–3815. DOI 10.1145/3832783.3834468

[47] Joosen A, Hassan A, Asenov M, Singh R, Darlow LN, Wang J, Deng Q, Barker A. 2025. Serverless Cold Starts and Where to Find Them. In: Proceedings of the Twentieth European Conference on Computer Systems (EuroSys 2025). New York: ACM, 938–953. DOI 10.1145/3689031.3696073

[48] Kehoe B, Patil S, Abbeel P, Goldberg K. 2015. A survey of research on cloud robotics and automation. IEEE Transactions on Automation Science and Engineering 12(2):398–409. DOI 10.1109/TASE.2014.2376492

[49] Kim J, Shin B, Chung J, Rhu M. 2026. The Cost of Dynamic Reasoning: Demystifying AI Agents and Test-Time Scaling from an AI Infrastructure Perspective. In: 2026 IEEE International Symposium on High-Performance Computer Architecture (HPCA). Piscataway: IEEE, 1–16. DOI 10.1109/hpca68181.2026.11408569

[50] Kitchenham B, Charters S. 2007. Guidelines for performing systematic literature reviews in software engineering. Technical Report EBSE-2007-01, Keele University and Durham University.

[51] Kothari N, Agarwal A, Joshi V, Kumar L. 2026. A Failure Taxonomy and Trace-Driven Debugging Toolkit for Reliable Agentic LLM Systems. In: 2026 International Conference on Artificial Intelligence, Systems, and Emerging Technologies (ICAISET). Piscataway: IEEE, 1–6. DOI 10.1109/icaiset66439.2026.11541708

[52] Li Z, Chen J, Jiao R, Zhao N, Wang Z, Zhang S, Wu Y, Jiang L, Yan L, Wang Z, Chen Z, Zhang W, Nie X, Sui K, Pei D. 2021b. Practical Root Cause Localization for Microservice Systems via Trace Analysis. In: Proceedings of the IEEE/ACM International Symposium on Quality of Service (IWQoS 2021). Piscataway: IEEE, 1–10. DOI 10.1109/iwqos52092.2021.9521340

[53] Li Q, Li L, Liu Z, Sun W, Li W, Li J, Zhao W. 2025. Cloud–Edge Collaboration for Industrial Internet of Things: Scalable Neurocomputing and Rolling-Horizon Optimization. IEEE Internet of Things Journal 12(12):19929–19943. DOI 10.1109/jiot.2025.3542428

[54] Li B, Peng X, Xiang Q, Wang H, Xie T, Sun J, Liu X. 2021a. Enjoy your observability: an industrial survey of microservice tracing and analysis. Empirical Software Engineering 27(1):25. DOI 10.1007/s10664-021-10063-9

[55] Li W, Tordsson J, Acker A, Kao O. 2020a. MicroRAS: Automatic Recovery in the Absence of Historical Failure Data for Microservice Systems. In: Proceedings of the IEEE/ACM International Conference on Utility and Cloud Computing (UCC 2020). Piscataway: IEEE, 227–236. DOI 10.1109/ucc48980.2020.00041

[56] Li W, Tordsson J, Elmroth E, Kao O. 2020b. MicroRCA: Root Cause Localization of Performance Issues in Microservices. In: Proceedings of the IEEE/IFIP Network Operations and Management Symposium (NOMS 2020). Piscataway: IEEE, 1–9. DOI 10.1109/noms47738.2020.9110353

[57] Liu W, Geng Y, Wang G, Liu X. 2022. A Fault-Tolerant and Real-Time Framework for Efficient Resilience in Edge Inference Systems. In: 2022 IEEE SmartWorld/UIC/ATC/ScalCom/DigitalTwin/PriComp/Metaverse Congress. Piscataway: IEEE, 121–128. DOI 10.1109/smartworld-uic-atc-scalcom-digitaltwin-pricomp-metaverse56740.2022.00043

[58] Liu D, He C, Peng X, Lin F, Zhang C, Gong S, Li Z, Ou J, Wu Z. 2021. MicroHECL: High-Efficient Root Cause Localization in Large-Scale Microservice Systems. In: Proceedings of the IEEE/ACM International Conference on Software Engineering: Software Engineering in Practice (ICSE-SEIP 2021). Piscataway: IEEE, 338–347. DOI 10.1109/icse-seip52600.2021.00043

[59] Liu X, Wen J, Chen Z, Li D, Chen J, Liu Y, Wang H, Jin X. 2023. FaaSLight: General Application-level Cold-start Latency Optimization for Function-as-a-Service in Serverless Computing. ACM Transactions on Software Engineering and Methodology 32(5):1–29. DOI 10.1145/3585007

[60] Long Z, Wu G, Chen X, Cui C, Chen W, Wei J. 2020. Fitness-guided Resilience Testing of Microservice-based Applications. In: Proceedings of the IEEE International Conference on Web Services (ICWS 2020). Piscataway: IEEE, 151–158. DOI 10.1109/icws49710.2020.00027

[61] Luo S, Lin C, Ye K, Xu G, Zhang L, Yang G, Xu H, Xu C. 2023. Optimizing Resource Management for Shared Microservices: A Scalable System Design. ACM Transactions on Computer Systems 42(1-2):1–28. DOI 10.1145/3631607

[62] Manner J, Endreb M, Heckel T, Wirtz G. 2018. Cold Start Influencing Factors in Function as a Service. In: Proceedings of the IEEE/ACM International Conference on Utility and Cloud Computing Companion (UCC Companion 2018). Piscataway: IEEE, 181–188. DOI 10.1109/ucc-companion.2018.00054

[63] Mittal A, Tadi G. 2025. Performance Optimization of LLM-Based Agentic Workloads in Kubernetes Environments: Bottlenecks, Root Causes, and Solutions. In: Proceedings of the International Conference on Computer and Applications (ICCA 2025). Piscataway: IEEE, 1–8. DOI 10.1109/icca66035.2025.11430741

[64] Molinari G, Ciravegna F. 2026. Towards Pervasive Distributed Agentic Generative AI–A State of The Art. ACM Computing Surveys 58(14):1–38. DOI 10.1145/3821566

[65] Nasiri N, Munshi N, Moser S, Pirvu M, Sundaresan V, Maier D, Premnath T, Böwing NC, Gopalakrishnan S, Shahrad M. 2026. In-Production Characterization of an Open Source Serverless Platform and New Scaling Strategies. In: Proceedings of the 21st European Conference on Computer Systems (EuroSys 2026). New York: ACM, 2054–2072. DOI 10.1145/3767295.3769377

[66] Page MJ, McKenzie JE, Bossuyt PM, Boutron I, Hoffmann TC, Mulrow CD, Shamseer L, Tetzlaff JM, Akl EA, Brennan SE, Chou R, Glanville J, Grimshaw JM, Hróbjartsson A, Lalu MM, Li T, Loder EW, Mayo-Wilson E, McDonald S, McGuinness LA, Stewart LA, Thomas J, Tricco AC, Welch VA, Whiting P, Moher D. 2021. The PRISMA 2020 statement: an updated guideline for reporting systematic reviews. BMJ 372:n71. DOI 10.1136/bmj.n71

[67] Petersen K, Vakkalanka S, Kuzniarz L. 2015. Guidelines for conducting systematic mapping studies in software engineering: an update. Information and Software Technology 64:1–18. DOI 10.1016/j.infsof.2015.03.007

[68] Priem J, Piwowar H, Orr R. 2022. OpenAlex: a fully-open index of scholarly works, authors, venues, institutions, and concepts. ArXiv preprint. arXiv:2205.01833

[69] Ren Y-Y, Lin X, Wang C, Chen L, Wang S, Gao K, Li D, Bai Y, Tian C, Zhang X, Li Y, Lin T, Zhai E. 2026. Networked Agent Memory and Causality Representation: Experiences towards Interpretable Cloud-Scale Root-Causing. In: Proceedings of the ACM SIGCOMM Conference (SIGCOMM 2026). New York: ACM, 787–808. DOI 10.1145/3789240.3829145

[70] Rodrigues H, Silva AR, Avritzer A. 2025. Assessment of performance and its scalability in microservice architectures: Systematic literature review. Journal of Systems and Software 230:112500. DOI 10.1016/j.jss.2025.112500

[71] Roy D, Zhang X, Bhave R, Bansal C, Las-Casas P, Fonseca R, Rajmohan S. 2024. Exploring LLM-Based Agents for Root Cause Analysis. In: Companion Proceedings of the 32nd ACM International Conference on the Foundations of Software Engineering (FSE Companion 2024). New York: ACM, 208–219. DOI 10.1145/3663529.3663841

[72] Safaryan G, Jindal A, Chadha M, Gerndt M. 2022. SLAM: SLO-Aware Memory Optimization for Serverless Applications. In: 2022 IEEE 15th International Conference on Cloud Computing (CLOUD). Piscataway: IEEE, 30–39. DOI 10.1109/cloud55607.2022.00019

[73] Satyanarayanan M. 2017. The emergence of edge computing. Computer 50(1):30–39. DOI 10.1109/MC.2017.9

[74] Satyanarayanan M, Klas G, Silva M, Mangiante S. 2019. The seminal role of edge-native applications. In: 2019 IEEE International Conference on Edge Computing (EDGE). Piscataway: IEEE, 33–40. DOI 10.1109/EDGE.2019.00022

[75] Schirmer T, Japke N, Greten S, Pfandzelter T, Bermbach D. 2023. The Night Shift: Understanding Performance Variability of Cloud Serverless Platforms. In: Proceedings of the 9th International Workshop on Serverless Computing (WoSC 2023). New York: ACM, 27–33. DOI 10.1145/3592533.3592808

[76] Schizas N, Karras A, Karras C, Sioutas S. 2022. TinyML for Ultra-Low Power AI and Large Scale IoT Deployments: A Systematic Review. Future Internet 14(12):363. DOI 10.3390/fi14120363

[77] Sebrechts M, Borny S, Wauters T, Volckaert B, De Turck F. 2021. Service Relationship Orchestration: Lessons Learned From Running Large Scale Smart City Platforms on Kubernetes. IEEE Access 9:133387–133401. DOI 10.1109/access.2021.3115438

[78] Shahrad M, Balkind J, Wentzlaff D. 2019. Architectural Implications of Function-as-a-Service Computing. In: Proceedings of the 52nd Annual IEEE/ACM International Symposium on Microarchitecture (MICRO 2019). New York: ACM. DOI 10.1145/3352460.3358296

[79] Sharma B, Nadig D. 2024. eBPF-Enhanced Complete Observability Solution for Cloud-native Microservices. In: ICC 2024 – IEEE International Conference on Communications. Piscataway: IEEE, 1980–1985. DOI 10.1109/icc51166.2024.10622329

[80] Shen C, Ouyang Q, Li F, Liu Z, Zhu L, Zou Y, Su Q, Yu T, Yi Y, Hu J, Zheng C, Wen B, Zheng H, Xu L, Pan S, Wu B, He X, Li Y, Tan J, Wang S, Pei D, Zhang W, Li F. 2023. Lindorm TSDB: A Cloud-Native Time-Series Database for Large-Scale Monitoring Systems. Proceedings of the VLDB Endowment 16(12):3715–3727. DOI 10.14778/3611540.3611559

[81] Shi W, Cao J, Zhang Q, Li Y, Xu L. 2016. Edge computing: vision and challenges. IEEE Internet of Things Journal 3(5):637–646. DOI 10.1109/JIOT.2016.2579198

[82] Shi Y, Yang Y, Yi C, Chen B, Cai J. 2024. Toward Online Reliability-Enhanced Microservice Deployment With Layer Sharing in Edge Computing. IEEE Internet of Things Journal 11(13):23370–23383. DOI 10.1109/jiot.2024.3385816

[83] Singh SK, Roy J. 2026. Scalable Explainability-as-a-Service (XaaS) for Edge AI Systems. In: SoutheastCon 2026. Piscataway: IEEE, 1–8. DOI 10.1109/southeastcon63549.2026.11476268

[84] Song S, Xu M, Zhang Z, Gao C, Zeng F, Ding Y, Ye K, Xu C. 2026. C-Koordinator: Interference-Aware Management for Large-Scale and Co-Located Microservice Clusters. Software Practice and Experience 56(5):506–524. DOI 10.1002/spe.70059

[85] Tran M-N, Vu XT, Kim Y. 2022. Proactive Stateful Fault-Tolerant System for Kubernetes Containerized Services. IEEE Access 10:102181–102194. DOI 10.1109/access.2022.3209257

[86] Tundo A, Mobilio M, Riganelli O, Mariani L. 2024. Monitoring Probe Deployment Patterns for Cloud-Native Applications: Definition and Empirical Assessment. IEEE Transactions on Services Computing 17(4):1636–1654. DOI 10.1109/tsc.2024.3349648

[87] Usman M, Ferlin S, Brunström A, Taheri J. 2022. A Survey on Observability of Distributed Edge & Container-Based Microservices. IEEE

Access 10:86904–86919. DOI 10.1109/access.2022.3193102

[88] Wang Z, Liu ZG, Zhang Y, Zhong A, Wang J, Yin F, Fan L, Wu L, Wen Q. 2024c. RCAgent: Cloud Root Cause Analysis by Autonomous Agents with Tool-Augmented Large Language Models. In: Proceedings of the ACM International Conference on Information and Knowledge Management (CIKM 2024). New York: ACM, 4966–4974. DOI 10.1145/3627673.3680016

[89] Wang L, Ma C, Feng X, Zhang Z, Yang H, Zhang J, Chen Z, Tang J, Chen X, Lin Y, Zhao WX, Wei Z, Wen J. 2024a. A survey on large language model based autonomous agents. Frontiers of Computer Science 18(6):186345. DOI 10.1007/s11704-024-40231-1

[90] Wang H, Ma W, Hu B, Li H, Si Q, Guo H, Huang B, Zhu Z, Zhang Y, Zhou Y, Zheng X. 2026a. DualLane: Fast and Reliable LLM Agents for Interactive AIOps via Dual-Path Planning. In: Proceedings of the ACM SIGKDD Conference on Knowledge Discovery and Data Mining (KDD 2026). New York: ACM, 8124–8134. DOI 10.1145/3770855.3818380

[91] Wang X, Wang B, Wu Y, Ning Z, Guo S, Yu FR. 2024b. A Survey on Trustworthy Edge Intelligence: From Security and Reliability to Transparency and Sustainability. IEEE Communications Surveys & Tutorials 27(3):1729–1757. DOI 10.1109/comst.2024.3446585

[92] Wang P, Xu J, Ma M, Lin W, Pan D, Wang Y, Chen P. 2018. CloudRanger: Root Cause Identification for Cloud Native Systems. In: Proceedings of the IEEE/ACM International Symposium on Cluster, Cloud and Grid Computing (CCGrid 2018). Piscataway: IEEE, 492–502. DOI 10.1109/ccgrid.2018.00076

[93] Wang J, Zhou X, Sun X, Zhang Y, Li Y, Wo T, Wang X, Hu C, Yang R. 2026b. Maestro: Workload-Aware Cross-Cluster Scheduling for LLM-based Multi-Agent Systems. In: Proceedings of the IEEE International Conference on Distributed Computing Systems (ICDCS 2026). Piscataway: IEEE, 250–260. DOI 10.1109/2575-8411.2026.00031

[94] Waseem M, Liang P, Shahin M, Di Salle A, Márquez G. 2021. Design, monitoring, and testing of microservices systems: The practitioners' perspective. Journal of Systems and Software 182:111061. DOI 10.1016/j.jss.2021.111061

[95] Wen Z, Chen Q, Deng Q, Niu Y, Song Z, Liu F. 2024. ComboFunc: Joint Resource Combination and Container Placement for Serverless Function Scaling With Heterogeneous Container. IEEE Transactions on Parallel and Distributed Systems 35(11):1989–2005. DOI 10.1109/tpds.2024.3454071

[96] Wen J, Chen Z, Sarro F, Wang S. 2025. Unveiling Overlooked Performance Variance in Serverless Computing. Empirical Software Engineering. DOI 10.1007/s10664-025-10615-3

[97] Wen Z, Lin T, Yang R, Ji S, Ranjan RK, Romanovsky AB, Lin C, Xu J. 2019. GA-Par: Dependable Microservice Orchestration Framework for Geo-Distributed Clouds. IEEE Transactions on Parallel and Distributed Systems 31(1):129–143. DOI 10.1109/tpds.2019.2929389

[98] Wen Z, Wang Y, Liu F. 2022. StepConf: SLO-Aware Dynamic Resource Configuration for Serverless Function Workflows. In: IEEE INFOCOM 2022 – IEEE Conference on Computer Communications. Piscataway: IEEE, 1868–1877. DOI 10.1109/infocom48880.2022.9796962

[99] Wohlin C. 2014. Guidelines for snowballing in systematic literature studies and a replication in software engineering. In: Proceedings of the 18th International Conference on Evaluation and Assessment in Software Engineering (EASE 2014). New York: ACM, Article 38. DOI 10.1145/2601248.2601268

[100] Wu C, Peng Q, Xia Y, Jin YJ, Hu Z. 2022. Towards cost-effective and robust AI microservice deployment in edge computing environments. Future Generation Computer Systems 141:129–142. DOI 10.1016/j.future.2022.10.015

[101] Xie Z, Zhang S, Geng Y, Zhang Y, Ma M, Nie X, Yao Z, Xu L, Sun Y, Li W, Pei D. 2024. Microservice Root Cause Analysis With Limited Observability Through Intervention Recognition in the Latent Space. In: Proceedings of the ACM SIGKDD Conference on Knowledge Discovery and Data Mining (KDD 2024). New York: ACM, 6049–6060. DOI 10.1145/3637528.3671530

[102] Xu Q, Gao Y, Wei J. 2024. An Empirical Study on Kubernetes Operator Bugs. In: Proceedings of the 33rd ACM SIGSOFT International Symposium on Software Testing and Analysis (ISSTA 2024). New York: ACM. DOI 10.1145/3650212.3680396

[103] Xu M, Song C, Ilager S, Gill SS, Zhao J, Ye K, Xu C. 2022. CoScal: Multifaceted Scaling of Microservices With Reinforcement Learning. IEEE Transactions on Network and Service Management 19(4):3995–4009. DOI 10.1109/tnsm.2022.3210211

[104] Xu M, Yang L, Wang Y, Gao C, Wen L, Xu G, Zhang L, Ye K, Xu C. 2023. Practice of Alibaba cloud on elastic resource provisioning for large-scale microservices cluster. Software Practice and Experience 54(1):39–57. DOI 10.1002/spe.3271

[105] Xue F, Ying T, Hu P, Wang L, Qing Z, Ding T. 2026. Fuzzing Microservice Resilience under Uncertainty with LLM-Guided Scenario Generation. In: Proceedings of the International Conference on Electronic Engineering and Informatics (EEI 2026). Piscataway: IEEE, 7–12. DOI 10.1109/eei70303.2026.11640489

[106] Xue Z, Zhao Y, Wang S, Chen K, Wang H. 2025. A Characterization Study of Bugs in LLM Agent Workflow Orchestration Frameworks. In: Proceedings of the IEEE/ACM International Conference on Automated Software Engineering (ASE 2025). Piscataway: IEEE, 3369–3380. DOI 10.1109/ase63991.2025.00278

[107] Yao S, Zhao J, Yu D, Du N, Shafran I, Narasimhan K, Cao Y. 2023. ReAct: synergizing reasoning and acting in language models. In: International Conference on Learning Representations (ICLR 2023).

[108] Yu G, Chen P, Li Y, Chen H, Li X, Zheng Z. 2023. Nezha: Interpretable Fine-Grained Root Causes Analysis for Microservices on Multi-modal Observability Data. In: Proceedings of the ACM Joint European Software Engineering Conference and Symposium on the Foundations of Software Engineering (ESEC/FSE 2023). New York: ACM, 553–565. DOI 10.1145/3611643.3616249

[109] Yu S, Wu H, Niu X, Nie C. 2025. A Systematic Literature Review on Fault Injection Testing of Microservice Systems. IEEE Transactions on Services Computing 18(6):4364–4385. DOI 10.1109/tsc.2025.3621564

[110] Zahid M, Pokorny FT. 2024. CloudGripper: An Open Source Cloud Robotics Testbed for Robotic Manipulation Research, Benchmarking and Data Collection at Scale. In: 2024 IEEE International Conference on Robotics and Automation (ICRA). Piscataway: IEEE, 12076–12082. DOI 10.1109/icra57147.2024.10611548

[111] Zayed A, Al-Said Ahmad A. 2024. Evaluating the Performance Resilience of Serverless Applications using Chaos Engineering. In: Proceedings of the IEEE International Conference on Software Quality, Reliability and Security Companion (QRS-C 2024). Piscataway: IEEE, 1316–1317. DOI 10.1109/qrs-c63300.2024.00172

[112] Zhang S, Jin P, Lin Z, Sun Y, Zhang B, Xia S, Li Z, Zhong Z, Ma M, Jin W, Zhang D, Zhu Z, Pei D. 2023. Robust Failure Diagnosis of Microservice System Through Multimodal Data. IEEE Transactions on Services Computing 16(6):3851–3864. DOI 10.1109/tsc.2023.3290018

[113] Zhang C, Peng X, Sha C, Zhang K, Fu Z, Wu X, Lin Q, Zhang D. 2022. DeepTraLog: trace-log combined microservice anomaly detection through graph-based deep learning. In: Proceedings of the 44th International Conference on Software Engineering (ICSE 2022). New York: ACM. DOI 10.1145/3510003.3510180

[114] Zhang S, Xia S, Fan W, Shi B, Xiong X, Zhong Z, Ma M, Sun Y, Pei D. 2025b. Failure Diagnosis in Microservice Systems: A Comprehensive Survey and Analysis. ACM Transactions on Software Engineering and Methodology 35(1):1–55. DOI 10.1145/3715005

[115] Zhang J, Yu X, Westerlund T. 2025a. Enhancing the Resilience of ROS 2-Based Multi-Robot Systems with Kubernetes: A Case Study on UWB-Based Relative Positioning. Sensors 25(16):5067. DOI 10.3390/s25165067

[116] Zhang X, Zhang H, Tan SH. 2026. Understanding Bugs in Modern Agentic Frameworks: A Study of Symptoms, Root Causes, and Triggering Conditions. In: Proceedings of the IEEE/ACM International Conference on Automated Software Engineering (ASE 2026). New York: ACM, 601–612. DOI 10.1145/3832783.3834378

[117] Zhao K, Zhou Z, Chen X, Zhou R, Zhang X, Yu S, Wu D. 2022. EdgeAdaptor: Online Configuration Adaption, Model Selection and Resource Provisioning for Edge DNN Inference Serving at Scale. IEEE Transactions on Mobile Computing 22(10):5870–5886. DOI 10.1109/tmc.2022.3189186

[118] Zheng M, Wu G, Zhang J, Wei J, Chen W, Zhu J, Liang Z. 2026. FaultWeave: Bounded Resilience Testing with Failure Diagnosis Capability for Microservice Applications. In: Proceedings of the 34th ACM International Conference on the Foundations of Software Engineering (FSE Companion 2026). New York: ACM, 857–867. DOI 10.1145/3803437.3805257

[119] Zhou Z, Chen X, Li E, Zeng L, Luo K, Zhang J. 2019a. Edge intelligence: paving the last mile of artificial intelligence with edge computing. Proceedings of the IEEE 107(8):1738–1762. DOI 10.1109/JPROC.2019.2918951

[120] Zhou H, Chen M, Lin Q, Wang Y, She X, Liu S, Gu R, Ooi BC, Yang J. 2018b. Overload Control for Scaling WeChat Microservices. In: Proceedings of the ACM Symposium on Cloud Computing (SoCC 2018). New York: ACM, 149–161. DOI 10.1145/3267809.3267823

[121] Zhou Z, Liu Z, Liu J, Shao Q, Wang Y, Shao K, Jin D, Xu F. 2026. ResMAS: Resilience Optimization in LLM-based Multi-agent Systems. Proceedings of the AAAI Conference on Artificial Intelligence 40(41):35176–35184. DOI 10.1609/aaai.v40i41.40824

[122] Zhou X, Peng X, Xie T, Sun J, Ji C, Li W, Ding D. 2018a. Fault Analysis and Debugging of Microservice Systems: Industrial Survey, Benchmark System, and Empirical Study. IEEE Transactions on Software Engineering 47(2):243–260. DOI 10.1109/tse.2018.2887384

[123] Zhou X, Peng X, Xie T, Sun J, Ji C, Liu D, Xiang Q, He C. 2019b. Latent Error Prediction and Fault Localization for Microservice Applications by Learning from System Trace Logs. In: Proceedings of the 27th ACM Joint European Software Engineering Conference and Symposium on the Foundations of Software Engineering (ESEC/FSE 2019). New York: ACM. DOI 10.1145/3338906.3338961

[124] Zhou Z, Zhang C, Ma L, Gu J, Qian H, Wen Q, Sun L, Li P, Tang Z. 2023. AHPA: Adaptive Horizontal Pod Autoscaling Systems on Alibaba Cloud Container Service for Kubernetes. Proceedings of the AAAI Conference on Artificial Intelligence 37(13):15621–15629. DOI 10.1609/aaai.v37i13.26852

**Table 1. Working definitions used in this review: cloud-native versus edge-native systems.**

| Dimension | Cloud-native systems | Edge-native systems |
|---|---|---|
| Design goal | Elastic, frequently changed services on shared data-centre infrastructure | Low latency, bandwidth saving, data locality and local autonomy close to devices and users |
| Typical infrastructure | Public, private or hybrid cloud data centres | Devices, gateways, base stations and micro data centres, usually with a cloud tier |
| Resource model | Abundant, homogeneous capacity provisioned on demand | Scarce, heterogeneous capacity; energy may be limited |
| Network assumption | Fast, mostly reliable data-centre network | Wireless or wide-area links with variable latency and bandwidth, and partitions |
| Dominant failure modes | Software faults, overload, cascading failures, faulty changes and misconfiguration | Node loss, disconnection, energy depletion and environmental effects, in addition to the cloud-native modes |
| Main reliability levers | Replication, rescheduling, overload control, automated recovery and rollback | Local fail-over, graceful degradation, offline operation, state reconciliation across tiers |
| Main scalability levers | Horizontal and vertical autoscaling, scale-to-zero in serverless platforms | Placement and offloading across tiers, adaptation of models and configurations to available capacity |
| Typical technologies | Microservices, containers, Kubernetes, service meshes, Function-as-a-Service (FaaS) | Lightweight container orchestration, edge AI runtimes, IoT and robot middleware |

Sources: CNCF (2018), Burns et al. (2016), Shi et al. (2016), Satyanarayanan (2017) and Satyanarayanan et al. (2019), complemented by the failure and scalability concerns reported in the reviewed studies.

**Table 2. Working definitions of the AI-related terms and the taxonomy categories that cover them.**

| Term | Definition used in this review | Category |
|---|---|---|
| Edge AI (edge intelligence) | Inference, and sometimes training, of AI models on edge nodes near the data source, often split with the cloud | C10 |
| On-device analytics | AI or analytics executed on the end device itself, including microcontroller-class TinyML | C10 |
| AI-driven edge intelligence | AI that manages edge resources: model and configuration selection, placement and offloading | C10; edge studies in C1, C6–C8 and C13 |
| Agentic AI for operations | LLM-based agents that detect, diagnose or mitigate failures of cloud systems | C11 |
| Reliability and scalability of agentic AI | Engineering agent applications, frameworks and their serving infrastructure so that they fail less, recover and scale | C12 |
| Physical AI (embodied AI) | AI that senses and acts in the physical world, such as robots, using edge or cloud resources | C13 |

Sources: Zhou et al. (2019a), Schizas et al. (2022), Yao et al. (2023), Wang et al. (2024a), Duan et al. (2022) and Kehoe et al. (2015). Learning-based resource management is classified by the mechanism it implements (for example, a learned autoscaler under C6); when the managed resources are at the edge, these studies are counted as AI-driven edge intelligence in Section 8.1.

**Table 3. Taxonomy of the 1,794 included primary studies with their evidence levels.**

| Category | Scope | RQ | n | EL1 | EL2 | EL3 | EL4 | EL? | EL3+EL4 (%) | Edge (%) |
|---|---|---|---|---|---|---|---|---|---|---|
| C1 Fault tolerance, high availability, recovery | Replication, fail-over, recovery, overload and failure-propagation control, reliability-aware deployment | RQ1 | 235 | 40 | 85 | 7 | 5 | 98 | 8.8 | 31.1 |
| C4 Resilience testing, chaos engineering | Fault injection, chaos experiments, resilience assessment | RQ1 | 35 | 1 | 19 | 0 | 1 | 14 | 4.8 | 17.1 |
| C5 Empirical failure studies | Characterisation of failures, incidents, bugs and practices | RQ1 | 11 | 0 | 1 | 6 | 1 | 3 | 87.5 | 0.0 |
| C2 Observability | Collection, storage and sampling of metrics, logs and traces; monitoring architectures | RQ2 | 157 | 3 | 67 | 4 | 5 | 78 | 11.4 | 15.3 |
| C3 Anomaly detection, root-cause analysis | Detecting, localising and classifying failures from telemetry | RQ2 | 351 | 20 | 120 | 25 | 6 | 180 | 18.1 | 9.1 |
| C6 Autoscaling, elasticity | Horizontal and vertical scaling, provisioning, resource configuration | RQ3 | 408 | 39 | 162 | 26 | 7 | 174 | 14.1 | 21.6 |
| C7 Cold start, performance variability | Serverless cold starts and run-to-run variability | RQ3 | 117 | 14 | 45 | 12 | 0 | 46 | 16.9 | 19.7 |
| C8 Scheduling, placement, orchestration | Placement, co-location and interference, service orchestration and configuration | RQ3 | 124 | 17 | 32 | 9 | 4 | 62 | 21.0 | 32.3 |
| C9 Performance evaluation, benchmarking | Scalability and performance measurement methods and comparisons | RQ3 | 55 | 1 | 35 | 0 | 1 | 18 | 2.7 | 7.3 |
| C10 Edge AI, on-device analytics | Reliability and scalability of AI inference and analytics on edge nodes and devices | RQ4 | 64 | 16 | 25 | 4 | 2 | 17 | 12.8 | 100.0 |
| C11 Agentic AI for operations | LLM agents that detect, diagnose or mitigate failures | RQ4 | 88 | 11 | 23 | 13 | 9 | 32 | 39.3 | 11.4 |
| C12 Reliability, scalability of agentic AI | Failures, testing, repair and serving of agent applications and frameworks | RQ4 | 77 | 10 | 25 | 5 | 5 | 32 | 22.2 | 31.2 |
| C13 Physical AI, edge–cloud | Cloud, fog and edge robotics; offloading for embodied AI | RQ4 | 72 | 9 | 25 | 1 | 0 | 37 | 2.9 | 95.8 |
| All | | | 1,794 | 181 | 664 | 112 | 46 | 791 | 15.8 | 25.5 |

EL1 simulation or model; EL2 prototype or testbed; EL3 production data; EL4 production deployment; EL? not determinable from the abstract. "EL3+EL4 (%)" is the share of production evidence among studies with a determinable level. "Edge (%)" is the keyword-derived share of studies addressing edge or edge–cloud settings (see Review Method). The study-level coding is in Dataset S1, sheet Included_studies.

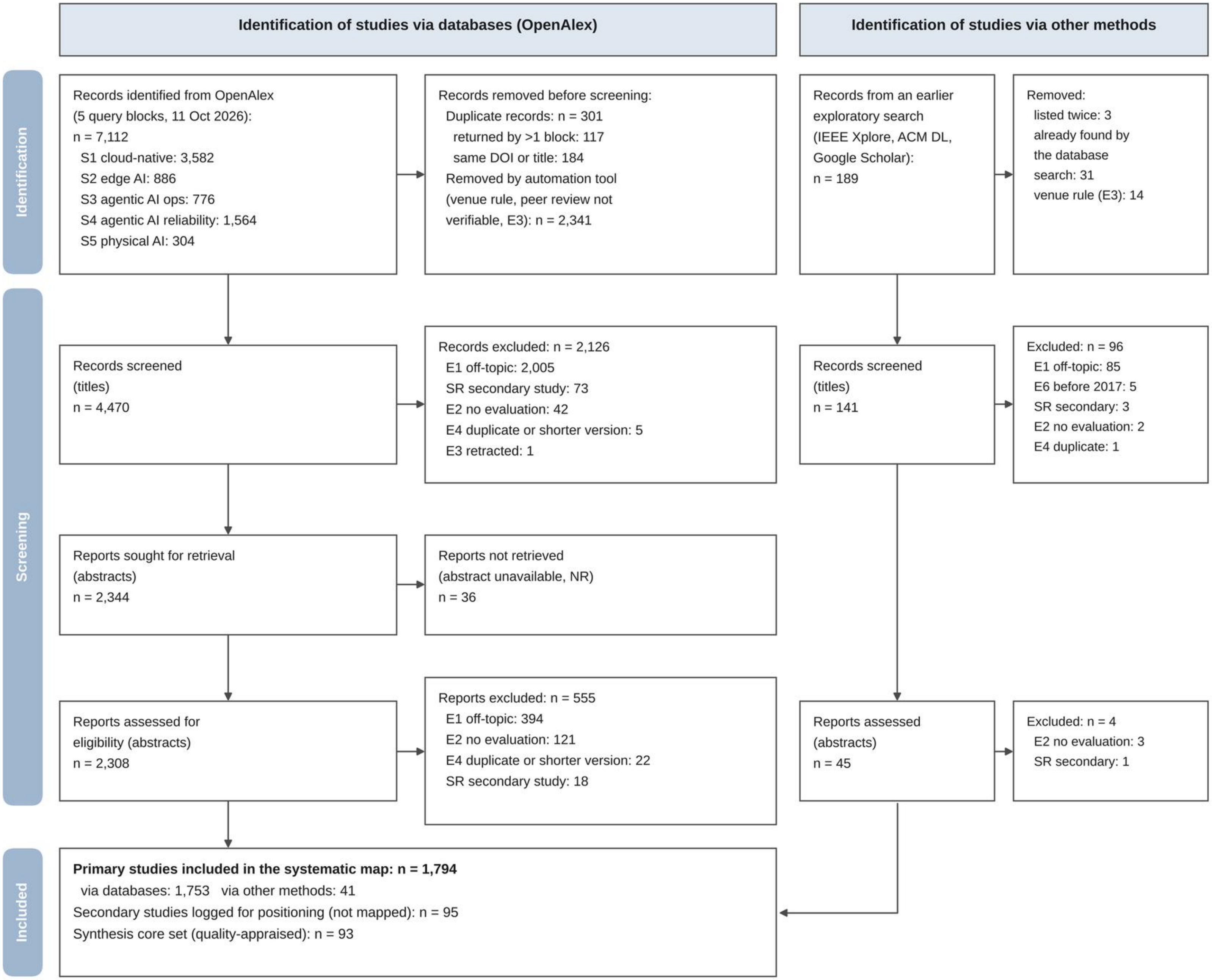
Identification of studies via databases (OpenAlex)
Identification of studies via other methods
Identification
Screening
Included
Records identified from OpenAlex (5 query blocks, 11 Oct 2026): n = 7,112
S1 cloud-native: 3,582
S2 edge AI: 886
S3 agentic AI ops: 776
S4 agentic AI reliability: 1,564
S5 physical AI: 304
Records removed before screening:
Duplicate records: n = 301
returned by >1 block: 117
same DOI or title: 184
Removed by automation tool (venue rule, peer review not verifiable, E3): n = 2,341
Records from an earlier exploratory search (IEEE Xplore, ACM DL, Google Scholar): n = 189
Removed:
listed twice: 3
already found by the database search: 31
venue rule (E3): 14
Records screened (titles) n = 4,470
Records excluded: n = 2,126
E1 off-topic: 2,005
SR secondary study: 73
E2 no evaluation: 42
E4 duplicate or shorter version: 5
E3 retracted: 1
Records screened (titles) n = 141
Excluded: n = 96
E1 off-topic: 85
E6 before 2017: 5
SR secondary: 3
E2 no evaluation: 2
E4 duplicate: 1
Reports sought for retrieval (abstracts) n = 2,344
Reports not retrieved (abstract unavailable, NR) n = 36
Reports assessed for eligibility (abstracts) n = 2,308
Reports excluded: n = 555
E1 off-topic: 394
E2 no evaluation: 121
E4 duplicate or shorter version: 22
SR secondary study: 18
Reports assessed (abstracts) n = 45
Excluded: n = 4
E2 no evaluation: 3
SR secondary: 1
Primary studies included in the systematic map: n = 1,794
via databases: 1,753 via other methods: 41
Secondary studies logged for positioning (not mapped): n = 95
Synthesis core set (quality-appraised): n = 93

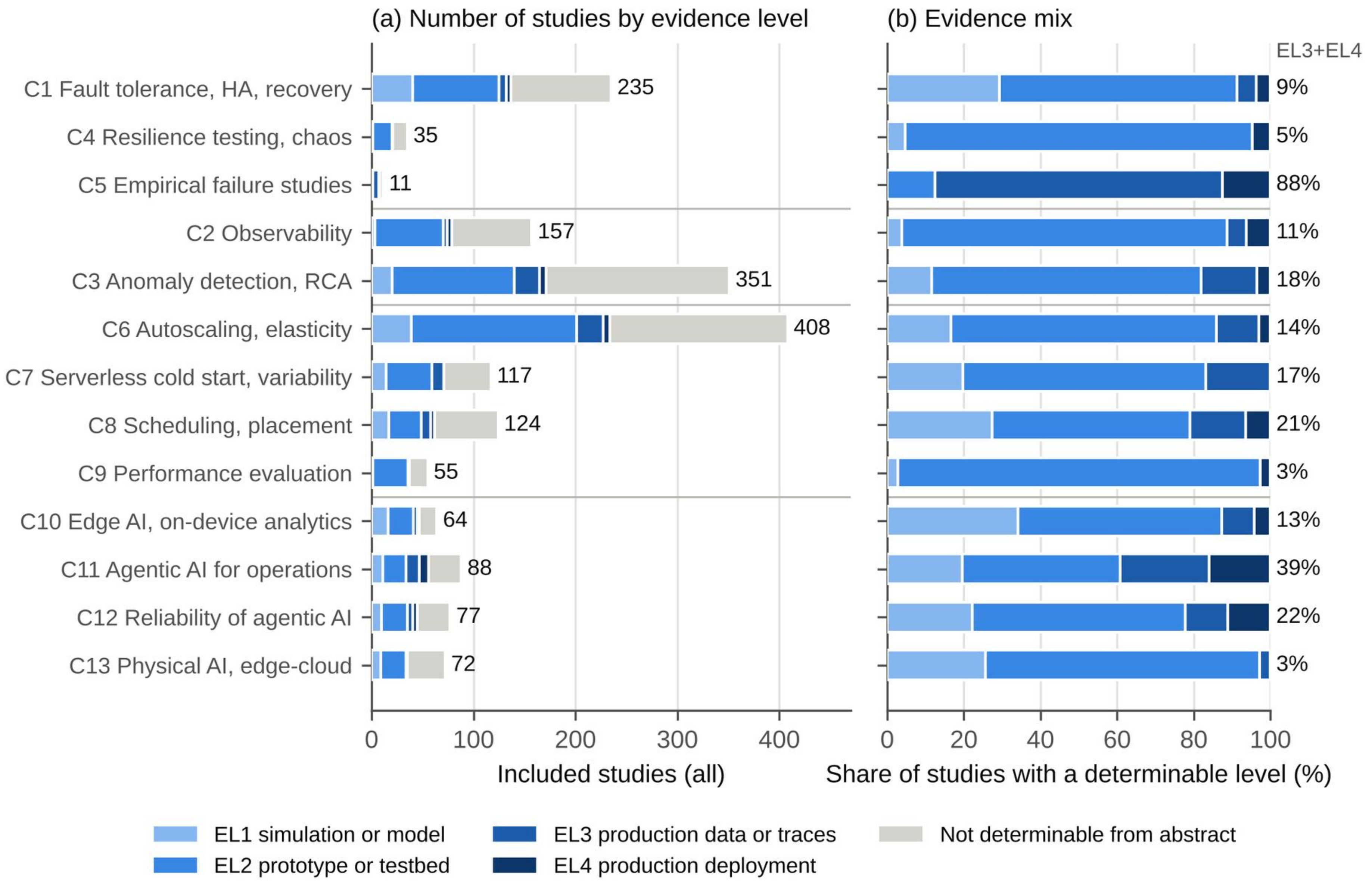

(a) Number of studies by evidence level
(b) Evidence mix
EL3+EL4
C1 Fault tolerance, HA, recovery
C4 Resilience testing, chaos
C5 Empirical failure studies
C2 Observability
C3 Anomaly detection, RCA
C6 Autoscaling, elasticity
C7 Serverless cold start, variability
C8 Scheduling, placement
C9 Performance evaluation
C10 Edge AI, on-device analytics
C11 Agentic AI for operations
C12 Reliability of agentic AI
C13 Physical AI, edge-cloud
235
35
11
157
351
408
117
124
55
64
88
77
72
9%
5%
88%
11%
18%
14%
17%
21%
3%
13%
39%
22%
3%
0
100
200
300
400
Included studies (all)
0
20
40
60
80
100
Share of studies with a determinable level (%)
EL1 simulation or model
EL2 prototype or testbed
EL3 production data or traces
EL4 production deployment
Not determinable from abstract

## Supplementary Table S1. Core studies for RQ1 (reliability and fault tolerance).

| Study | Mechanism or focus | Evaluation setting | EL | Key reported result | Abstract QA |
|---|---|---|---|---|---|
| Chabbi et al. (2026) | Static analysis of error propagation to find cascading-failure risks | Uber, more than 6,000 microservices | EL4 | >3,800 risks found, >1,300 remediated; ~5% false positives | 4.0 |
| Zhou et al. (2018b) | Collaborative load shedding for overload control | WeChat backend, five years | EL4 | High success rate and fairness under overload | 3.5 |
| Wen et al. (2019) | Dependability-aware orchestration across geo-distributed clouds | Large real-world datasets | EL3 | +42.34% over a greedy security-aware method; ~4× faster solver | 5.5 |
| He et al. (2024) | Detecting changes that reduce fault resilience | 256 incidents from four systems; three datasets | EL3 | 37.87% of erroneous changes reduce resilience; F1 0.9 | 5.5 |
| Li et al. (2020a) | Recovery-action selection without failure history | Kubernetes testbed, injected anomalies | EL2 | 94.7% of faulty services recovered; ≥44.3% less interference | 5.5 |
| Tran et al. (2022) | Fault prediction with stateful pod migration | Kubernetes testbed | EL2 | Fewer QoS violations than default Kubernetes | 4.5 |
| Bian et al. (2024) | Asynchronous shared log for stateful serverless workflows | Serverless runtime prototype | EL2 | Up to −45.8% latency and −46.5% logging overhead | 5.0 |
| Shi et al. (2024) | Reliability-aware edge deployment with container-layer sharing | Simulation | EL1 | +12.4% reliability; −28.57% total delay | 5.0 |
| Gao et al. (2023) | Coverage-guided crash and reboot injection | ZooKeeper, HDFS, HBase | EL2 | Four new and one known crash-recovery bug | 5.0 |
| Long et al. (2020) | Fitness-guided resilience testing reusing integration tests | Microservice benchmark | EL2 | Exposes recovery-logic bugs within a fixed time budget | 4.5 |
| Xue et al. (2026) | LLM-generated multi-stage uncertainty campaigns | DeathStarBench, Online Boutique | EL2 | +56% failure paths; 86% SLO-violation coverage vs 62% and 48% | 6.0 |
| Zheng et al. (2026) | Bounded fault-combination testing with minimal failure sets | Industrial application, 512 microservices, three months | EL4 | 237 resilience vulnerabilities found | 4.5 |
| Chen et al. (2026) | Multi-tool chaos framework for cloud–edge Kubernetes | 11,965 fault-injection scenarios | EL2 | Cloud–edge ~80% more stable under delay and partition; cloud ~47% more resilient under bandwidth limits | 5.5 |
| Zayed & Al-Said Ahmad (2024) | Delay-injection chaos experiments for serverless functions | AWS Lambda | EL2 | Performance impact against a fault-free baseline | 4.0 |
| Zhou et al. (2018a) | Industrial survey and benchmark with replicated fault cases | TrainTicket; 22 industrial fault cases | EL2 | Tracing and visualisation improve debugging | 5.0 |
| Waseem et al. (2021) | Practitioner survey and interviews | 106 responses; six interviews | EL3 | Monitoring and testing lack dedicated solutions | 5.0 |
| Ghosh et al. (2022) | Study of production incidents | Large-scale cloud service | EL3 | Delays in detection, root-causing and mitigation | 4.0 |
| de Toledo et al. (2021) | Repaying architectural technical debt | ~1,000 microservices, financial company | EL4 | Incidents −84%; critical and high-priority incidents about −90% | 5.0 |
| Ezaz et al. (2024) | Critical-path analysis of performance variability | Alibaba traces | EL3 | Locates interactions with high variability | 4.0 |
| Xu et al. (2024) | Study of Kubernetes operator bugs | 210 bugs in 36 operators | EL3 | Root causes, impacts and fixes characterised | 5.5 |

EL: evidence level (Supplementary Table S8). QA: abstract-based reporting score out of 6 (Supplementary Table S9). Results are quoted from each study and are not comparable across studies, which use different systems, workloads and metrics.

## Supplementary Table S2. Core studies for RQ2 (observability, anomaly detection and root-cause analysis).

| Study | Mechanism or focus | Evaluation setting | EL | Key reported result | Abstract QA |
|---|---|---|---|---|---|
| Li et al. (2021a) | Industrial survey of tracing and trace analysis | Interviews in ten companies | EL3 | Tracing pipelines common; machine learning seldom used | 4.5 |
| Ashok et al. (2024) | Trace reconstruction without instrumentation | Benchmarks and a production dataset | EL3 | ~90% reconstruction accuracy | 4.5 |
| Sharma & Nadig (2024) | Kernel-level (eBPF) observability | Chameleon bare-metal testbed | EL2 | Up to 210× lower CPU overhead; up to 159% less memory than common agents | 5.5 |
| Tundo et al. (2024) | Monitoring-probe deployment patterns | VMs and containers | EL2 | Targets <1% CPU; probe holders up to 10 GiB memory | 5.5 |
| Shen et al. (2023) | Time-series database for monitoring at scale | Different data scales | EL2 | Outperforms two open-source TSDBs; stable during scaling | 4.5 |
| He et al. (2023) | Observability-preserving trace sampling | Benchmarks and production systems | EL2 | Context only | – |
| Li et al. (2020b) | Attributed-graph root-cause localisation | Kubernetes benchmark, injected anomalies | EL2 | 89% precision; 97% mean average precision | 5.5 |
| Wang et al. (2018) | Causal impact graph without topology | IBM Bluemix incidents; simulation | EL3 | +10% accuracy over earlier approaches | 5.5 |
| Liu et al. (2021) | Anomaly-propagation chains with pruning | Alibaba production | EL4 | Top-3 hit ratio 68%; localisation 30 → 5 min | 5.0 |
| Li et al. (2021b) | Unsupervised trace-based localisation | Open benchmark, production system, bank | EL4 | Top-1 accuracy +44.8% over unsupervised methods | 6.0 |
| Gan et al. (2019) | Deep learning to anticipate QoS violations | Local clusters and larger deployments | EL2 | 91% of violations anticipated; 84% avoided | 4.5 |
| Zhang et al. (2023) | Multimodal diagnosis with a graph neural network | Real-world datasets | EL3 | Localisation +20.9–368%; failure type +11.0–169% | 5.5 |
| Yu et al. (2023) | Event-pattern comparison over multimodal data | Two microservice applications | EL2 | Average top-1 accuracy 89.77% | 5.5 |
| Zhang et al. (2022) | Trace–log graph with deep one-class learning | Microservice benchmark | EL2 | Precision 0.93; recall 0.97 | 6.0 |
| Xie et al. (2024) | Latent-variable RCA under limited observability | eBay production; testbed | EL4 | Top-1 recall +9.7–57.9% | 6.0 |
| Zhou et al. (2019b) | Error prediction and fault localisation from trace logs | Fault-injection data | EL? | Context only | – |

EL: evidence level (Supplementary Table S8). QA: abstract-based reporting score out of 6 (Supplementary Table S9); "–" marks studies cited for context only. Results are not comparable across studies.

## Supplementary Table S3. Core studies for RQ3 (scalability and resource management).

| Study | Mechanism or focus | Evaluation setting | EL | Key reported result | Abstract QA |
|---|---|---|---|---|---|
| Zhou et al. (2023) | Forecast- and model-based pod autoscaling | Alibaba Cloud Container Service, since 2021 | EL4 | CPU usage +10%; resource cost −20% or more | 5.0 |
| Xu et al. (2023) | Proactive and reactive provisioning for microservice clusters | Alibaba clusters | EL4 | Resource usage +10–15% at required latency | 5.5 |
| Gias et al. (2019) | Layered queueing model for autoscaling | Microservice testbed | EL2 | ~30–37% higher throughput than rule-based control | 5.0 |
| Abdullah et al. (2020) | Burst-aware predictive autoscaling | Trace-driven simulation | EL1 | ×5.17 fewer SLO violations at higher cost | 5.5 |
| Xu et al. (2022) | Reinforcement learning over horizontal, vertical scaling and brownout | Sock Shop, Robot Shop | EL2 | Response time −19–29%; successful transactions +6–10% | 6.0 |
| Al Qassem et al. (2023) | Random-forest proactive autoscaler | Open-source prototyping platform | EL2 | Utilisation up to 90%; p95 latency improved up to 95% | 6.0 |
| Fang et al. (2026) | Hierarchical graph learning for autoscaling | Real-world request traces | EL3 | Average response time up to −80.16% under a VM budget | 5.5 |
| Wen et al. (2022) | Per-step memory configuration of function workflows | AWS Lambda | EL2 | Up to 40.9% cost saving within SLOs | 5.5 |
| Nasiri et al. (2026) | In-production characterisation of a serverless platform | Production platform | EL3 | Context only | 2.5 |
| Joosen et al. (2025) | Characterisation of cold starts | Huawei trace: 85 billion requests, 11.9 million cold starts | EL3 | Cold starts up to 7 s or 3 s, with different dominant components per region | 5.0 |
| Manner et al. (2018) | Factors influencing cold starts | AWS Lambda, Azure Functions; 49,500 executions | EL2 | User-perceived overhead 300 ms to 24 s | 5.5 |
| Liu et al. (2023) | Application-level code-loading optimisation | AWS Lambda, Google Cloud Functions | EL2 | Response latency up to −42.05% (−19.21% on average) | 5.5 |
| Cai et al. (2024) | Priority-based scheduling against cold starts | Production traces | EL3 | 1.4× faster than native; −23% and −14.8% vs two systems | 5.5 |
| Chen (2025) | Snapshots and pre-warming for serverless LLM inference | Azure Functions trace | EL3 | Cold start up to 49× faster; GPU idle cost −32% | 6.0 |
| Schirmer et al. (2023) | Long-term performance variability | Google Cloud Functions, months | EL2 | Diurnal differences up to 15%; 3× more cold starts at week start | 4.5 |
| Wen et al. (2025) | Run-to-run performance variance | Serverless functions | EL? | Context only | – |
| Song et al. (2026) | Interference-aware co-location | Alibaba co-located clusters | EL4 | Interference prediction >90.3%; P50–P99 latency −16.7–36.1% | 6.0 |
| Luo et al. (2023) | Latency-target scaling for shared microservices | Benchmarks; trace-driven simulation | EL2 | 5× lower SLA-violation probability; 1.6× less resources | 5.5 |
| Wen et al. (2024) | Heterogeneous containers with in-place vertical scaling | Real traces on a local cluster | EL3 | Up to 3.01× faster scaling; cost up to −42.6% | 5.5 |
| Bhasi et al. (2024) | Heterogeneous hardware and GPU sharing for serverless inference | 16 workloads, real traces, six nodes | EL3 | SLO compliance up to +13.3%; cost up to −86% | 5.5 |
| Sebrechts et al. (2021) | Service relationships as Kubernetes metadata | Smart-city platforms; proof of concept | EL4 | 0.44 s overhead per service for change propagation | 4.0 |
| Safaryan et al. (2022) | Trace-based memory configuration of serverless applications | AWS Lambda, four applications | EL2 | >95% of requests within SLO | 4.5 |
| Henning & Hasselbring (2022) | Configurable scalability benchmarking method | Two public clouds, one private cloud | EL2 | ≤5 repetitions of ≤5 minutes suffice | 5.5 |
| Henning & Hasselbring (2023) | Scalability of stream-processing frameworks as microservices | >740 h; up to 110 instances, 1M messages/s | EL2 | Approximately linear scalability; framework choice changes resource cost | 5.5 |

| Study | Mechanism or focus | Evaluation setting | EL | Key reported result | Abstract QA |
|---|---|---|---|---|---|
| Blinowski et al. (2022) | Monolith versus microservices | Local, Azure Spring Cloud, Azure App Service | EL2 | Monolith faster on one machine; scaling out degrades beyond a point | 5.0 |
| Copik et al. (2021) | Serverless benchmark suite | Several cloud platforms | EL2 | Context only | – |
| Shahrad et al. (2019) | Architectural implications of FaaS | FaaS platform characterisation | EL2 | Context only | – |
| Eismann et al. (2022) | Stability of serverless performance tests | Repeated tests over ten months | EL? | Context only | – |

EL: evidence level (Supplementary Table S8). QA: abstract-based reporting score out of 6 (Supplementary Table S9); "–" marks studies cited for context only. Results are not comparable across studies.

## Supplementary Table S4. Core studies for RQ4 (edge AI, agentic AI and physical AI).

| Study | Mechanism or focus | Evaluation setting | EL | Key reported result | Abstract QA |
|---|---|---|---|---|---|
| Liu et al. (2022) | Coded-computation fault tolerance for edge inference | Real-world datasets | EL2 | Max–median latency gap up to 2.5× smaller | 5.0 |
| Wu et al. (2022) | Robust placement of multi-model edge AI microservices | Real-world edge environments | EL2 | Lower cost and QoS loss under node failures | 5.0 |
| Zhao et al. (2022) | Online model, configuration and resource adaptation for edge inference | Trace-driven simulation | EL1 | Empirical competitive ratio ≤1.4 | 5.0 |
| Li et al. (2025) | Scalable network variants with rolling-horizon scheduling | Simulation and prototype | EL2 | Feasibility under delay constraints | 3.5 |
| Singh & Roy (2026) | Explainability as a service for edge AI | Three real-world deployments | EL4 | Latency −38% | 5.0 |
| Hajder et al. (2026) | Post-mortem of a field edge AI system | Wildlife-deterrence deployment | EL4 | Energy collapse within 24 h; event-driven redesign | 4.0 |
| Jiang et al. (2026) | Causal skeleton with memory-augmented multi-agent RCA | Kuaishou production, >6 months | EL4 | AC@1 0.88 and 0.79; diagnosis time −77.3% | 5.0 |
| Ren et al. (2026) | Multi-agent root-causing with networked memory | Global cloud network, 12 months, >3,000 incidents | EL4 | Root-causing time −25.8%; accuracy 94.6% | 5.0 |
| Fu et al. (2025b) | Multi-agent on-call automation | Video service production, two months | EL4 | Response 21 s; triage 4 s on average | 5.0 |
| Wang et al. (2026a) | Dual-path planning for interactive AIOps agents | Alibaba Cloud ECS, >1 year | EL4 | Accuracy 96.5%; median latency 4.2 s; agent error rate 7.1% | 4.0 |
| Wang et al. (2024c) | Tool-augmented autonomous RCA agent | Alibaba Cloud production data | EL3 | Outperforms ReAct; integrated into a production platform | 4.0 |
| Roy et al. (2024) | ReAct agent with retrieval tools for RCA | Production incidents of a large IT company | EL3 | Competitive with baselines; higher factual accuracy | 4.5 |
| Chen et al. (2025) | Multi-agent SRE with transactional no-regression | AIOpsLab, ITBench | EL2 | ≥1.5× mitigation success over earlier agents | 6.0 |
| de Castro et al. (2026) | Closed-loop benchmark of remediation agents | Live 5G core on Kubernetes; 900 experiments | EL2 | Tool access raises success from 10.9% to 61.1% | 6.0 |
| Xue et al. (2025) | Bugs in LLM agent workflow frameworks | 1,026 bugs in three frameworks | EL3 | Nine root causes; six symptom categories | 5.5 |
| Zhang et al. (2026) | Bugs in multi-agent frameworks | 409 bugs in five frameworks | EL3 | Agent-specific root causes; tests in 47% of model-layer fixes | 5.5 |
| Al Hasan & Biswas (2026) | Operational safety failures of coding agents | 547 incidents from 16,586 issues | EL3 | 326 high or critical; >65% in bug fixing and setup | 5.5 |
| Huang et al. (2026) | Tool-invocation reliability diagnostics | 1,980 deterministic tests; edge hardware | EL2 | Tool initialisation dominates failures; 96.6% success at 7.3 s | 5.5 |
| Kothari et al. (2026) | Failure taxonomy, tracing and auto-repair for agents | 300 tasks and a WebArena subset | EL2 | Success 52.4% → 68.9%; <5% overhead | 6.0 |
| Zhou et al. (2026) | Resilient topologies and prompts for multi-agent systems | Multiple tasks and models | EL2 | Higher resilience under perturbations | 4.0 |
| Kim et al. (2026) | System-level cost of agent reasoning | Representative agents | EL2 | Diminishing returns; rising latency variance and power | 5.0 |
| Wang et al. (2026b) | Workload-aware scheduling of multi-agent LLM workflows | Prototype; trace-driven simulation | EL2 | KV-cache HBM −67.2%; SLO attainment +23.6 pp | 5.5 |
| Mittal & Tadi (2025) | Optimising agent workloads on Kubernetes | Production deployments | EL4 | 2–5× throughput; 33× faster cold starts | 3.5 |
| Chinchali et al. (2021) | Learned offloading policy for cloud robotics | Simulation and hardware | EL2 | 1.3–2.3× better vision performance | 5.5 |
| Huang et al. (2022) | Edge-accelerated multi-robot SLAM | Evaluation and field prototype | EL2 | Processing latency up to −39.31% | 5.0 |

| Study | Mechanism or focus | Evaluation setting | EL | Key reported result | Abstract QA |
|---|---|---|---|---|---|
| Ichnowski et al. (2023) | Cloud and fog robotics platform for ROS 2 | Common robot applications | EL2 | SLAM latency −50%; motion planning 45× faster | 6.0 |
| Zhang et al. (2025a) | Kubernetes orchestration for resilient ROS 2 robots | Jetson edge cluster, induced failures | EL2 | Positioning error under failure scenarios | 4.0 |
| Alirezazadeh et al. (2022) | Algorithm allocation across robots, fog and cloud | Real-world data | EL3 | Improves on an earlier allocation method | 4.5 |
| Zahid & Pokorny (2024) | Open cloud robotics testbed | 32 work cells, 10 Gbit/s | EL2 | Dataset of >100 h and ~4 million images | 5.0 |

EL: evidence level (Supplementary Table S8). QA: abstract-based reporting score out of 6 (Supplementary Table S9). Results are not comparable across studies.

**Supplementary Table S5. Related secondary studies and their overlap with this review.**

| Study | Type | Scope | Overlap |
|---|---|---|---|
| Zhang et al. (2025b) | Survey (98 papers) | Failure diagnosis in microservice systems: concepts, datasets, toolkits, metrics | RQ2 |
| Fu et al. (2025a) | Survey | Root-cause localisation in microservice systems and the role of AI techniques | RQ2 |
| Usman et al. (2022) | Survey | Observability of distributed edge and container-based microservices | RQ2 |
| Gomes et al. (2025) | Systematic mapping study | Observability of microservice-based applications | RQ2 |
| Yu et al. (2025) | Systematic literature review | Fault injection testing of microservice systems | RQ1 |
| Golec et al. (2024) | Systematic review | Cold-start latency in serverless computing | RQ3 |
| Jeong et al. (2025) | Survey | Autoscaling techniques in cloud-native computing | RQ3 |
| Rodrigues et al. (2025) | Systematic literature review (29 of 801 studies) | Assessment of performance and scalability in microservice architectures | RQ3 |
| Wang et al. (2024b) | Survey | Trustworthy edge intelligence, including reliability | RQ4 |
| Schizas et al. (2022) | Systematic review | TinyML for low-power AI and IoT deployments | RQ4 |
| Molinari & Ciravegna (2026) | Survey | Distributed agentic generative AI in pervasive computing | RQ4 |
| Afrin et al. (2021) | Survey | Resource allocation and service provisioning in multi-agent cloud robotics | RQ4 |
| This review | Systematic mapping study with evidence-graded synthesis | Reliability and scalability across cloud-native, edge-native, edge AI, agentic AI and physical AI | RQ1–RQ4 |

## Supplementary Table S6. Search blocks run in OpenAlex on 11 October 2026.

| Block | Purpose | Fields searched (per concept group) | Coverage | Records |
|---|---|---|---|---|
| S1 | Cloud-native reliability, scalability and observability | Title AND title | 2017–2026 | 3,582 |
| S2 | Edge AI and on-device analytics | Title and abstract AND title | 2017–2026 | 886 |
| S3 | Agentic AI for cloud operations | Title AND title and abstract AND title and abstract | 2023–2026 | 776 |
| S4 | Reliability and scalability of agentic AI applications | Title AND title AND title and abstract | 2023–2026 | 1,564 |
| S5 | Physical AI on the edge–cloud continuum | Title and abstract (three groups) | 2017–2026 | 304 |
| Total | | | | 7,112 |

Document types: article, conference paper, book chapter, review. The exact strings follow in the text; Dataset S1 (sheet Search_log) records the API filters and the hit count at retrieval for each block.

**Supplementary Table S7. Five-point relevance scale and how it was applied.**

| Level | Definition | Decision | Example from the corpus |
|---|---|---|---|
| 5 | The main contribution is a reliability, scalability, observability or diagnosis mechanism (or an empirical analysis of one) for a target system of this review, evaluated quantitatively | Include | Root-cause localisation of performance issues in Kubernetes microservices (Li et al., 2020b) |
| 4 | The main contribution addresses an RQ topic, but in an adjacent setting (general distributed or cloud infrastructure) or with a mainly qualitative evaluation | Include | A cloud-native time-series database for large-scale monitoring (Shen et al., 2023) |
| 3 | An RQ topic is a secondary focus, for example a performance or platform study with a measured effect on availability, latency or scalability | Include | A cloud and fog robotics platform that reduces latency and network use (Ichnowski et al., 2023) |
| 2 | The target systems or qualities are mentioned only in passing | Exclude (E1) | — |
| 1 | Unrelated to the research questions | Exclude (E1) | — |

**Supplementary Table S8. Evidence levels used to separate simulation, prototype and production evidence.**

| Level | Evaluation setting | Typical examples |
|---|---|---|
| EL1 | Analytical model or simulation, including trace-driven simulation | Queueing model or simulator of an autoscaling policy |
| EL2 | Laboratory prototype or benchmark testbed, including public-cloud experiments with synthetic or benchmark workloads | Fault injection on an open-source microservice benchmark; experiments on AWS Lambda |
| EL3 | Real production data: replayed production traces, incident reports, bug repositories or practitioner surveys | Replay of Alibaba, Azure or Huawei traces; analysis of GitHub issues or incident reports |
| EL4 | Deployment or evaluation in a production environment with reported field results | A mechanism running in an operator's production system |
| EL? | Not determinable from the abstract | — |

**Supplementary Table S9. Abstract-based reporting-appraisal items (each scored 1, 0.5 or 0).**

| Item | Question |
|---|---|
| QA1 | Are the problem and the objective clearly stated? |
| QA2 | Is the mechanism, or for empirical studies the study method, described well enough to be understood? |
| QA3 | Is the evaluation setting described (system, workload, dataset)? |
| QA4 | Is the mechanism compared with a baseline or an alternative? |
| QA5 | Are quantitative results reported? |
| QA6 | Are the artifact or data available, or is a public benchmark used? |

**Supplementary Table S10. Synthesis core set: evidence levels and abstract-based reporting scores by category.**

| Category | Core studies | EL1 | EL2 | EL3 | EL4 | EL? | Scored | Mean abstract QA | Abstract QA ≥ 4.5 | Abstract QA 3–4 | Context only |
|---|---|---|---|---|---|---|---|---|---|---|---|
| C1 | 8 | 1 | 3 | 2 | 2 | 0 | 8 | 4.81 | 6 | 2 | 0 |
| C4 | 6 | 0 | 5 | 0 | 1 | 0 | 6 | 4.92 | 5 | 1 | 0 |
| C5 | 6 | 0 | 1 | 4 | 1 | 0 | 6 | 4.75 | 4 | 2 | 0 |
| C2 | 6 | 0 | 4 | 2 | 0 | 0 | 5 | 4.90 | 5 | 0 | 1 |
| C3 | 10 | 0 | 4 | 2 | 3 | 1 | 9 | 5.50 | 9 | 0 | 1 |
| C6 | 9 | 1 | 4 | 2 | 2 | 0 | 9 | 5.17 | 8 | 0 | 1 |
| C7 | 7 | 0 | 3 | 3 | 0 | 1 | 6 | 5.33 | 6 | 0 | 1 |
| C8 | 6 | 0 | 2 | 2 | 2 | 0 | 6 | 5.17 | 5 | 1 | 0 |
| C9 | 6 | 0 | 5 | 0 | 0 | 1 | 3 | 5.33 | 3 | 0 | 3 |
| C10 | 6 | 1 | 3 | 0 | 2 | 0 | 6 | 4.58 | 4 | 2 | 0 |
| C11 | 8 | 0 | 2 | 2 | 4 | 0 | 8 | 4.94 | 6 | 2 | 0 |
| C12 | 9 | 0 | 5 | 3 | 1 | 0 | 9 | 5.11 | 7 | 2 | 0 |
| C13 | 6 | 0 | 5 | 1 | 0 | 0 | 6 | 5.00 | 5 | 1 | 0 |
| All | 93 | 3 | 46 | 23 | 18 | 3 | 87 | 5.05 | 73 | 13 | 7 |

Abstract-based QA scores range from 0 to 6 (Supplementary Table S9). "Context only" marks studies without a complete abstract-based score or scoring below 3; these studies are cited but do not support findings on their own. Study-level scores and blank author-only full-text-check fields are in Dataset S1, sheet Core_set_QA.

## Supplementary Table S11. Main conclusions, their supporting evidence and the research directions they motivate.

| Conclusion | Core-set support (evidence level) | Map-level support | Research direction |
|---|---|---|---|
| Failure propagation can be contained, and root causes localised, in large production microservice systems | Chabbi et al. (2026), Zhou et al. (2018b), Liu et al. (2021), Li et al. (2021b), Xie et al. (2024) (all EL4) | C3: 18.1% production evidence | Independent replication across operators; shared incident datasets with ground truth, as released by Xie et al. (2024) |
| Guided fault-space exploration finds more resilience failures than random or rule-based injection | Xue et al. (2026), Gao et al. (2023) (EL2); Zheng et al. (2026) (EL4) | C4: 4.8% production evidence | Open resilience benchmarks with fault libraries for microservice, serverless and edge settings |
| Predictive and model-based autoscaling outperforms threshold rules; production savings of 10–20% at one operator | Zhou et al. (2023), Xu et al. (2023) (EL4); Gias et al. (2019), Xu et al. (2022), Al Qassem et al. (2023) (EL2) | C6: 14.1% production evidence | Evaluate learned autoscalers on public production traces against common baselines, reporting cost and SLO compliance jointly |
| Cold-start latency varies widely, and mitigations work in testbeds and trace replays | Joosen et al. (2025), Cai et al. (2024), Chen (2025) (EL3); Liu et al. (2023), Manner et al. (2018) (EL2) | C7: no study coded EL4 | Production evaluations of cold-start mitigations; experiment designs that control for run-to-run and diurnal variability (Schirmer et al., 2023) |
| Reliability mechanisms for edge settings rest mostly on models and simulations | Shi et al. (2024) (EL1); Chen et al. (2026) (EL2); Hajder et al. (2026) (EL4) | Edge studies: 26.9% simulation vs 14.0% for cloud-only; no edge study in C5 | Empirical failure studies of edge deployments; testbeds with energy, partition and bandwidth faults |
| LLM agents for operations are used in production, and their reliability rests on tool grounding, structural constraints and safety properties | Jiang et al. (2026), Ren et al. (2026), Fu et al. (2025b) (EL4); Chen et al. (2025), de Castro et al. (2026) (EL2) | C11: 39.3% production evidence | Live, shared benchmarks for agentic operations with safety metrics; reporting of agent-induced errors; independent replication |
| Agentic applications fail in agent-specific ways, and reliability practices exist mainly as research prototypes | Xue et al. (2025), Zhang et al. (2026), Al Hasan & Biswas (2026) (EL3); Kothari et al. (2026), Wang et al. (2026b) (EL2) | C12: 22.2% production evidence | Dependability engineering for agent frameworks: testing of model-integration layers, transactional semantics for agent actions, SLO-aware agent serving |
| Offloading to edge and fog resources scales physical AI, but its reliability is rarely the main research question | Chinchali et al. (2021), Huang et al. (2022), Ichnowski et al. (2023) (EL2) | C13: 2.9% production evidence; no study at relevance level 5 | Reliability-centred studies of cloud and fog robotics, including safe behaviour on disconnection |

All conclusions are limited to the 1,794 studies documented in Dataset S1. Absence of evidence in this corpus is not evidence that a mechanism is ineffective or untested elsewhere.

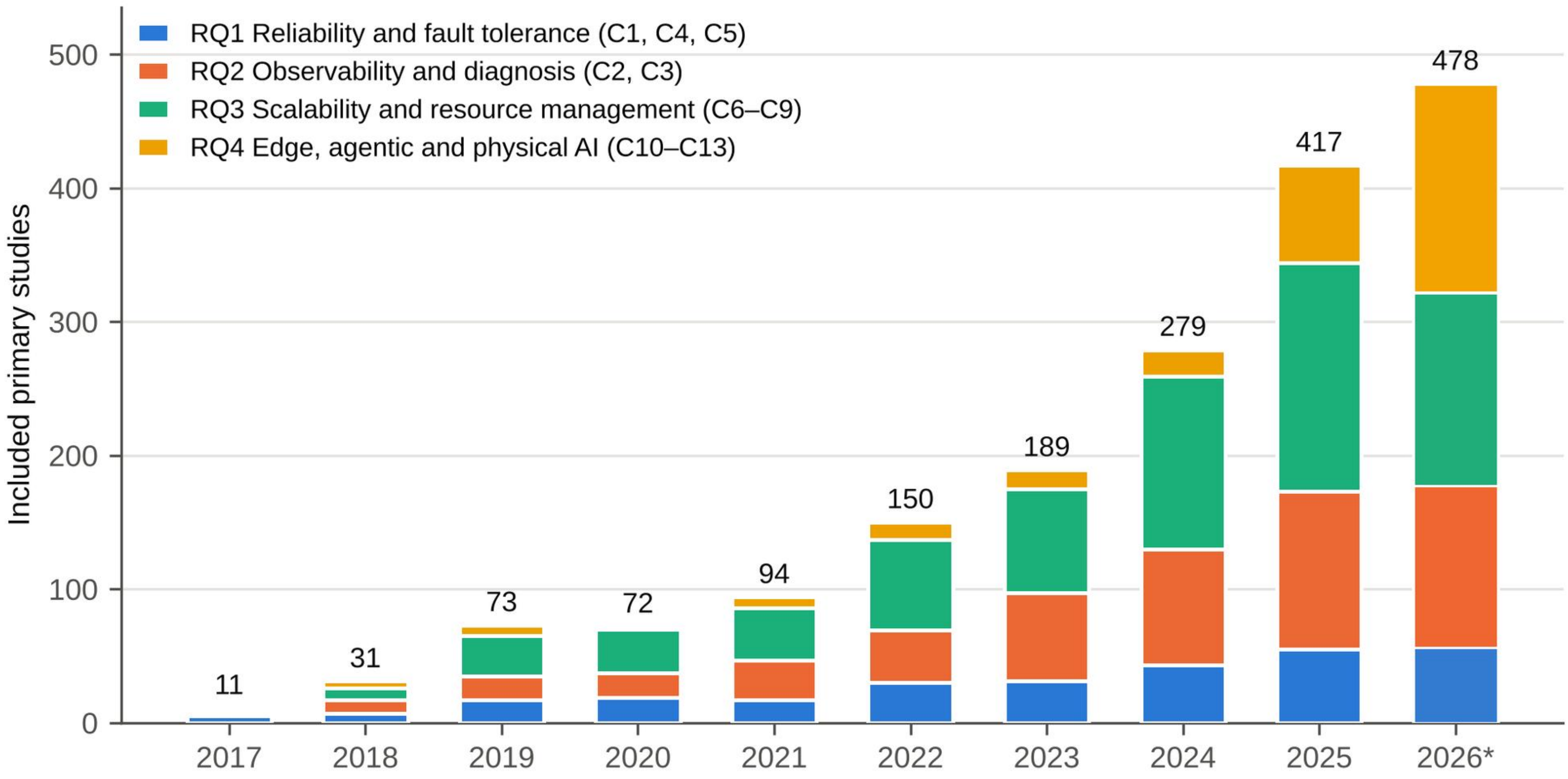

* 2026 covers publications indexed up to the search date (11 October 2026).